\RequirePackage{fix-cm}
\documentclass[smallextended]{svjour3}       %
\smartqed  %
\usepackage{graphicx}
\usepackage{natbib}
\usepackage{rotating}
\usepackage{paralist}
\usepackage{booktabs}
\usepackage{tabularx}
\usepackage{multirow}
\usepackage[table]{xcolor}
\usepackage{hyperref}
\usepackage{siunitx}
\usepackage{hyperref}
\usepackage{amsmath}
\DeclareMathOperator*{\argmin}{argmin}
\usepackage{framed}

\makeatletter
\let\MYcaption\@makecaption
\makeatother
\usepackage[font=footnotesize,labelformat=simple]{subcaption}
\makeatletter
\let\@makecaption\MYcaption
\makeatother

\journalname{Empirical Software Engineering}
\begin{document}

\title{Can Offline Testing of Deep Neural Networks Replace Their Online Testing?\thanks{This work has received funding from Luxembourg’s National Research Fund (FNR) under grant BRIDGES2020/IS/14711346/FUNTASY, the European Research Council under the European Union’s Horizon 2020 research and innovation programme (grant agreement No 694277), IEE S.A. Luxembourg, and NSERC of Canada under the Discovery and CRC programs.
Donghwan Shin was partially supported by the Basic Science Research Programme through the National Research Foundation of Korea (NRF) funded by the Ministry of Education (2019R1A6A3A03033444).}}
\subtitle{A Case Study of Automated Driving Systems}

\author{
Fitash Ul Haq\and
Donghwan Shin\and
Shiva Nejati\and
Lionel Briand
}

\institute{
Fitash Ul Haq \and Donghwan Shin \and Shiva Nejati \and Lionel Briand
\at SnT, University of Luxembourg \\
\email{\{fitash.ulhaq, donghwan.shin, shiva.nejati, lionel.briand\}@uni.lu}
\and
Shiva Nejati \and Lionel Briand
\at University of Ottawa \\
\email{\{snejati, lbriand\}@uottawa.ca}
}

\date{Received: date / Accepted: date}

\maketitle

\begin{abstract}
We distinguish two general modes of testing for Deep Neural Networks (DNNs):
Offline testing where DNNs are tested as individual units based on test datasets
obtained without involving the DNNs under test, and online testing where
DNNs are embedded into a specific application environment and tested in a
closed-loop mode in interaction with the application environment. Typically,
DNNs are subjected to both types of testing during their development life cycle
where offline testing is applied immediately after DNN training and online
testing follows after offline testing and once a DNN is deployed within a
specific application environment. In this paper, we study the relationship
between offline and online testing. Our goal is to determine \emph{how offline
testing and online testing  differ or complement one another} and \emph{if
offline testing results can be used to help reduce the cost of online testing?}
Though these questions are generally relevant to all autonomous systems, we
study them in the context of automated driving systems where, as study subjects,
we use DNNs automating end-to-end controls of steering functions of self-driving
vehicles. Our results show that offline testing is less effective than
online testing as many safety violations identified by online testing could not
be identified by offline testing, while large prediction errors generated by
offline testing always led to severe safety violations detectable by online
testing. Further, we cannot exploit offline testing results to
reduce the cost of online testing in practice since we are not able to
identify specific situations where offline testing could be as accurate as
online testing in identifying safety requirement violations.
\end{abstract}

\section{Introduction}
\label{sec:intro}

Deep Neural Networks (DNNs) have been widely adopted in many real-world
applications, such as image classification~\citep{ciresan2012}, natural language
processing~\citep{SutskeverVL14}, and speech recognition~\citep{DengHK13}.
Recent successes of DNNs on such practical problems make them key enablers of smart
and autonomous systems such as automated-driving vehicles. As DNNs are
increasingly used in safety critical autonomous systems, the challenge of
ensuring safety and reliability of DNN-based systems emerges as a fundamental
software verification problem.

\begin{figure}
	\centering
	\includegraphics[width=0.9\linewidth]{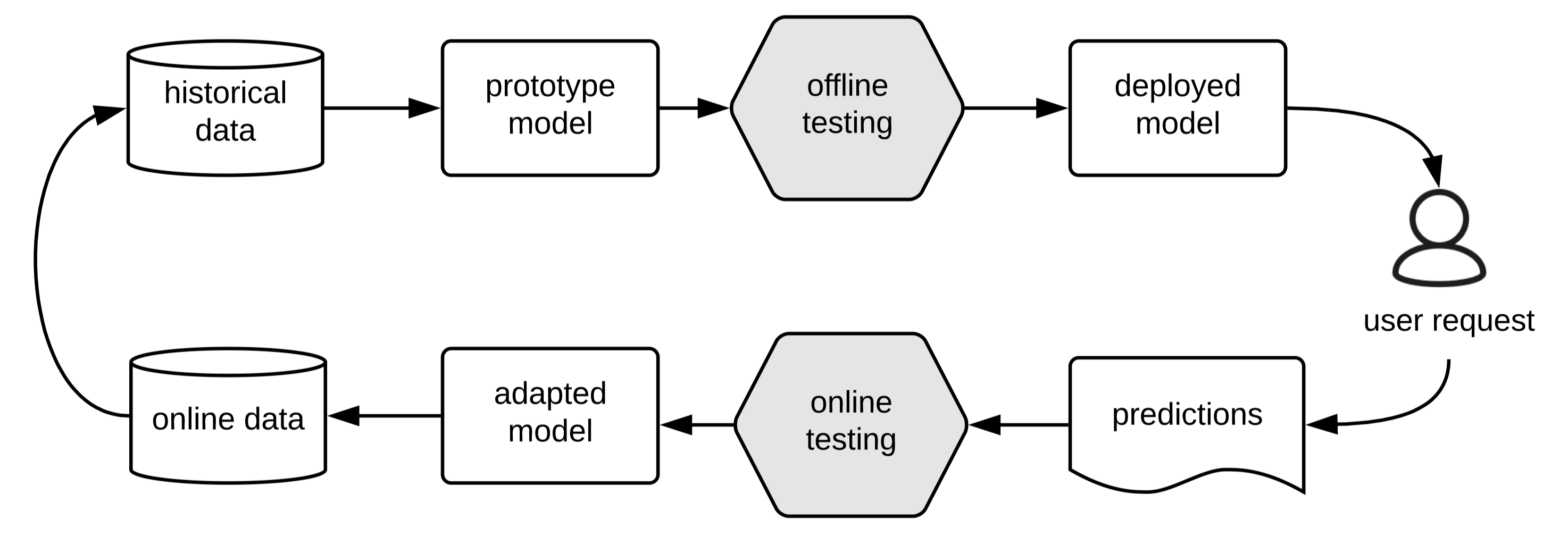}
	\caption{Idealized Workflow of ML testing~\citep{9000651}}
	\label{fig:workflow-ml-testing}
\end{figure}

A main distinction between DNN testing (or in general, testing Machine Learning
components) and traditional software testing is that the process of DNN testing
follows a specific workflow that involves two testing phases, i.e., \emph{offline
testing} and \emph{online testing}, as shown in Figure~\ref{fig:workflow-ml-testing}.
Offline testing is a necessary and standard step in developing Machine
Learning (ML) models and is applied immediately after training a DNN model.
It is used to ensure that the trained DNN model is
sufficiently accurate when applied to new data (i.e., test data). Online testing,
in contrast, is performed after deploying a DNN into a specific application
(e.g., an automated driving system) and evaluates DNN interactions with the
application environment and users. Specifically, in offline testing, DNNs
are tested as a unit in an open-loop mode. They are fed with test inputs
generated without involving the DNN under test, either manually or automatically.
The outputs of DNNs are then typically evaluated by assessing their prediction
error, which is the difference between the expected test outputs (i.e., test
oracles) and the outputs generated by the DNN under test. In online testing,
however, DNNs are tested within an application environment in a closed-loop
mode. They receive test inputs generated by the environment, and their outputs
are, then, directly fed back into the environment. Online testing evaluates DNNs
by monitoring the requirements violations they trigger, for example related to safety.
Given the safety critical nature of many systems relying on DNNs (e.g., self-driving cars),
most online testing approaches rely on simulators, as testing DNNs embedded into a real and
operational environment is expensive, time consuming and often dangerous.

Recently, many DNN testing techniques and algorithms have been proposed in the literature~\citep{9000651}.
However, the majority of the existing DNN testing techniques are
either specifically designed for offline testing or even if they could be
applied in an online setting, they are still solely evaluated and assessed in an
offline setting. This is partly because offline testing matches the standard
checking of an ML model in terms of prediction accuracy, that does not require
the DNN to be embedded into an application environment and can be readily
carried out with either manually generated or automatically generated test data.
Given the increasing availability of open-source data, a large part of offline
testing research uses open-source, manually-generated real-life test data.
Online testing, on the other hand, necessitates embedding a DNN into an
application environment, either real or simulated.

Even though online testing is less studied, it remains an important
phase for DNN testing for a number of reasons~\citep{9000651}. First, the test
data used for offline testing may not be representative of the data that a DNN
should eventually be able to handle when it is embedded into a specific
application. Second, in contrast to offline testing, online testing is able to
assess the DNN interactions with an application environment and can reveal failures that
can only occur with real applied scenarios (e.g., if an accident actually happens in
self-driving cars on real driving scenarios).
In other words, while offline testing results are limited to assessing
prediction errors or prediction accuracy, online testing results can be
used to directly assess system-level requirements (e.g., whether or not an
accident happened, or if there is a security breach, or if there is a data loss
or communication error).

At a high-level, we expect offline testing to be faster and less expensive than
online testing because offline testing does not require a closed-loop
environment to generate test inputs. However, there is limited insight as to how
these two testing modes compare with one another
with respect to their ability to reveal faulty behaviors and
most particularly those leading to safety violations.
This, for example, would depend if large prediction errors identified by offline
testing always lead to safety violations detectable by online testing, or if the
safety violations identified by online testing translate into large prediction
errors. Answers to these questions would enable us to better understand
the limitations of the two testing stages and their relationship.

We investigated the above questions in an empirical study and presented the results in a
conference paper~\citep{ICST20} published in International Conference on
Software Testing, Verification and Validation (ICST 2020). Though the
investigated questions are generally relevant to all autonomous systems, we
performed an empirical study to compare offline testing and online testing in the
context of Automated Driving Systems (ADS). In particular, our study aimed to
ultimately answer the following research question: \emph{How do offline and
online testing results differ and complement each other?} To answer this
question, we used open-source DNN models developed to automate steering
functions of self-driving vehicles~\citep{udacity:challenge}. To enable online
testing of these DNNs, we integrated them into a powerful, high-fidelity
physics-based simulator of self-driving cars~\citep{prescan}. The simulator
allows us to specify and execute scenarios capturing various road traffic
situations, different pedestrian-to-vehicle and vehicle-to-vehicle interactions,
and different road topologies, weather conditions and infrastructures. As a
result, in our study offline and online testing approaches were compared with
respect to the data generated automatically using a simulator. To ensure that
this aspect does not impact the validity of our comparison, we investigated the
following research question as a pre-requisite of the above question: \emph{Can we
use simulator-generated data as a reliable substitute to real-world data
for the purpose of DNN testing?}

While the above research questions provide insights on the relationship between
offline and online testing results, it is still unclear how we can use offline
and online testing together in practice such that we can minimize cost
and maximize the effectiveness of testing DNNs. As the ML testing workflow in
Figure~\ref{fig:workflow-ml-testing} suggests, offline testing always precedes
online testing and given that offline testing is considerably less expensive
than online testing, it is beneficial if we can exploit offline testing results to
reduce the cost of online testing by running fewer tests. In this article, we introduce a new
research question to determine \emph{if offline testing results can be used
to help reduce the cost of online testing?} Our goal is to identify whether we can
characterize the test scenarios (conditions) where offline and online testing
results are the same with high probability. To do so, we propose a novel
heuristic approach to infer such conditions from limited number of offline
and online testing data in an efficient and effective way.

The contributions of this article are summarized below:
\begin{enumerate}
\item We show that we can use simulator-generated datasets in lieu of real-life
datasets for testing DNNs in our application context. Our comparison between
online and offline testing using such datasets show that offline and online
testing results frequently differ, and specifically, offline testing results are
often not able to find faulty behaviors due to the lack of error accumulation over time. As a
result, many safety violations identified by online testing could not be
identified by offline testing as they did not cause large prediction errors.
However, all the large prediction errors generated by offline testing led to
severe safety violations detectable by online testing.

\item We provide a three-step approach to infer (learn) conditions
characterizing agreement and disagreement between offline and online testing
results while minimizing the amount of the data required to infer the conditions
and maximizing the statistical confidence of the results.

\item We were not able to infer any conditions that can characterize agreement
between offline and online testing results with a probability higher than 71\%.
This means that, in general, we cannot exploit offline testing results to
reduce the cost of online testing in practice.
\end{enumerate}

The first contribution is mainly the result of our first two research questions
presented in our earlier work~\citep{ICST20}. In this article, we have, however,
extended our first contribution in two ways: \emph{First,} our earlier work used
two ADS DNNs from the Udacity challenge~\citep{udacity:challenge}, namely Autumn
and Chauffeur. In this article, we add another DNN model (Komanda) from
the same Udacity challenge
to the set of our study subjects to strengthen our results. We also
surveyed other ADS-DNNs from the Udacity challenge and other sources, but we were
not able to find any other suitable study subject candidate since other
public ADS-DNNs are either significantly more inaccurate (higher prediction errors)
than our three selected DNNs or their inputs and outputs were not compatible
with our simulator, and hence, we could not test them in an online setting.
\emph{Second,} in this article, we provide additional correlation analysis
between offline and online testing to support our results.

The second and third contributions are completely new and have not been
presented before. In addition, in this article, we refine and extend ideas from
our previous work and extend our discussion of the related literature.
Through our experiments, we collected both offline and online testing results for more than
700 test scenarios in total, taking around 350 hours of simulations,
resulting in around 50 GB of simulator-generated images.
To facilitate the replication of our study, we have made all the experimental
materials, including simulator-generated data, publicly available~\citep{supp}.

The rest of the paper is organized as follows: Section~\ref{sec:background}
provides background on DNNs for autonomous vehicles, introduces offline and
online testing, describes our proposed domain model that is used to configure
simulation scenarios for automated driving systems, and formalizes the main
concepts in offline and online testing used in our experiments.
Section~\ref{sec:expr} reports on the empirical evaluation.
Section~\ref{sec:offline-online} surveys the existing research on online and
offline testing for automated driving systems. Section~\ref{sec:conclusion}
concludes the paper.

\section{Background}\label{sec:background}
This section provides the basic concepts that will be used throughout the article.

\subsection{DNNs in ADS}\label{sec:ADS-DNN}

\begin{figure}
	\centering
	\includegraphics[width=0.7\linewidth]{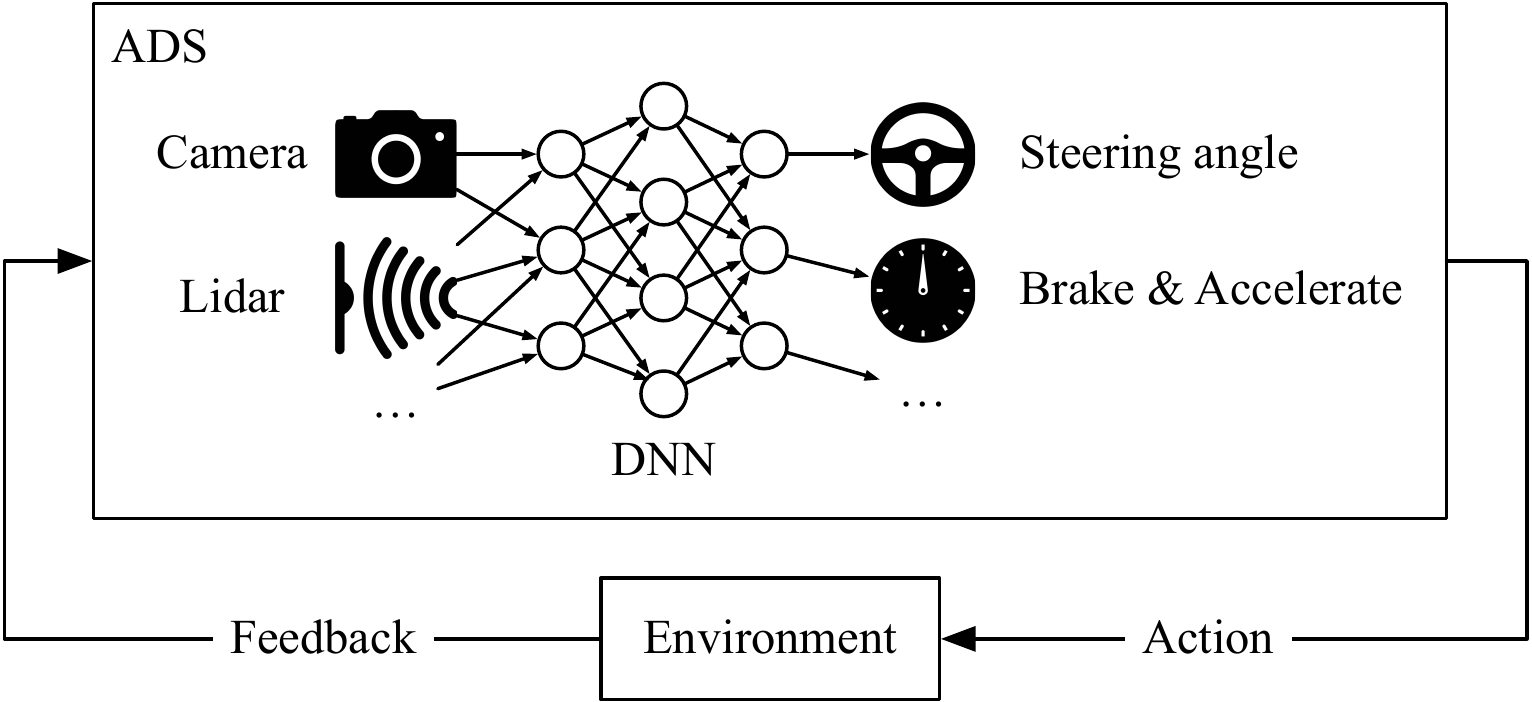}
	\caption{Overview of DNN-based ADS}
	\label{fig:dnn-in-ads}
\end{figure}

Depending on the ADS design, DNNs may be used in two ways to automate the
driving task of a vehicle: One design approach is to incorporate DNNs into the
ADS perception layer, primarily to do \emph{semantic
segmentation}~\citep{6248074}, i.e., to classify and label each and every pixel
in a given image. The ADS software controller then decides what commands should
be issued to the vehicle's actuators based on the classification results
produced by the DNN~\citep{pomerleau1989alvinn}. An alternative design approach
is to use DNNs to perform the \emph{end-to-end} control of a
vehicle~\citep{udacity:challenge} (e.g., Figure~\ref{fig:dnn-in-ads}). In this
case, DNNs directly generate the commands to be sent to the vehicle's actuators
after processing images received from cameras. Our approach to compare offline
and online testing of DNNs is applicable to both ADS designs. In the comparison
provided in this article, however, we use DNN models automating the end-to-end
control of the steering function since these models are publicly available
online and have been extensively used in recent studies on DNN
testing~\citep{DeepTest,DeepRoad,DeepGauge,Surprise}. In particular, we
investigate the DNN models from the Udacity self-driving challenge as our study
subjects~\citep{udacity:challenge}. We refer to this class of DNNs as ADS-DNNs
in the remainder of the article. Specifically, an ADS-DNN receives as input
images from a front-facing  camera mounted on a vehicle, and generates a
steering angle command for the vehicle.

\subsection{Test Data Sources}\label{sec:data-sources}
We identify two sources for generating test data for testing ADS-DNNs: (1)
real-life driving and (2) driving simulator.

For our ADS-DNN models, a \emph{real-life dataset} is a video or a sequence of
images captured by a camera mounted on a vehicle's dashboard while the vehicle
is being driven by a human driver. The steering angle of the vehicle applied by
the human driver is recorded for the duration of the video and each image
(frame) of the video in this sequence is labelled by its corresponding steering
angle. This yields a sequence of manually labelled images to be used for testing
DNNs. There are, however, some drawbacks with test datasets captured from
real-life~\citep{KALRA2016182}. Specifically, data generation is expensive, time
consuming and lacks diversity. The latter issue is particularly critical since
driving scenes, driving habits, as well as objects, infrastructures and roads in
driving scenes, can vary widely across countries, continents, climates, seasons,
day times, and even drivers.

Another source of test data generation is to use simulators to automatically
generate videos capturing various driving scenarios. There are increasingly more
high-fidelity and advanced physics-based simulators for self-driving vehicles
fostered by the needs of the automotive industry, which increasingly relies on
simulators to improve their testing and verification practices. There are
several examples of commercial ADS simulators (e.g., PreScan~\citep{prescan} and
Pro-SiVIC~\citep{prosivic}) and a number of open source ones (e.g.,
CARLA~\citep{carla} and LGSVL~\citep{9294422}). These simulators incorporate
dynamic models of vehicles (including vehicles' actuators, sensors and cameras)
and humans as well as various environment aspects (e.g., weather conditions,
different road types, different infrastructures). The simulators are highly
configurable and can be used to generate desired driving scenarios. In our work,
we use the PreScan simulator to generate test data for ADS-DNNs. PreScan is a
widely-used, high-fidelity commercial ADS simulator in the automotive domain and
has been used by our industrial partner. In Section~\ref{sec:domain}, we present
the domain model that define the inputs used to configure the simulator, and
describe how we automatically generate scenarios that can be used to test
ADS-DNNs. Similar to real-life videos, the videos generated by our simulator are
sequences of labelled images such that each image is labelled by a steering
angle. In contrast to real-life videos, the steering angles generated by the
simulator are automatically computed based on the road trajectory as opposed to
being generated by a human driver.

The simulator-generated test datasets are cheaper and faster to produce compared
to real-life ones. In addition, depending on how advanced and comprehensive the
simulator is, we can achieve a higher-level of diversity in the
simulator-generated datasets by controlling and varying the objects, roads,
weather, and other various features. However, it is not yet clear whether
simulator-generated images can be used in lieu of real images since the latter
may have higher resolution, showing more natural texture, and look more
realistic. In this article, we conduct an empirical study in
Section~\ref{sec:expr} to investigate \emph{if we can use simulator-generated
images as a reliable alternative to real images for testing ADS-DNNs}.

\subsection{Domain Model}\label{sec:domain}

\begin{figure}
	\centering
	\includegraphics[width=0.9\linewidth]{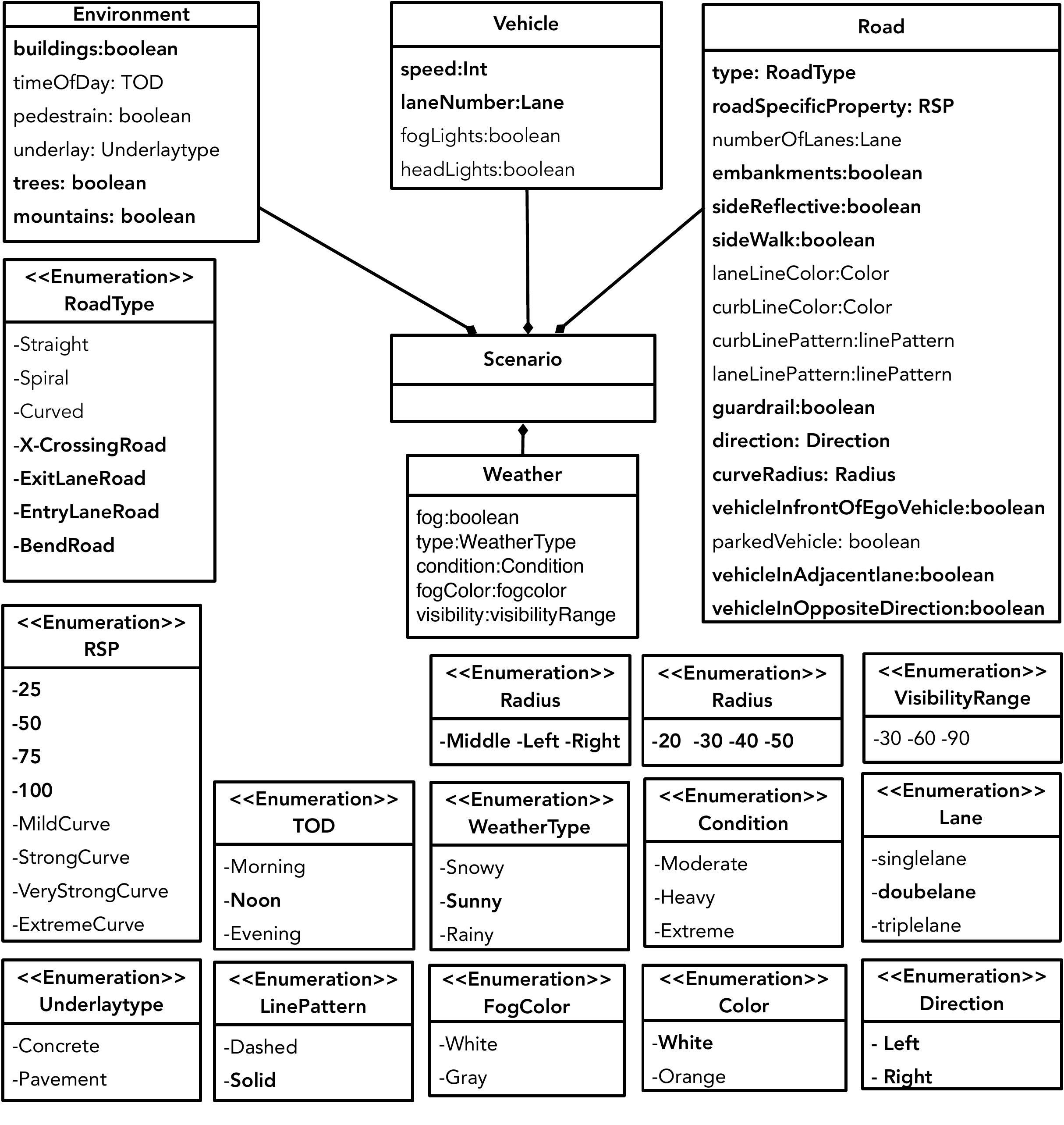}
	\caption{Complete domain model for scenario generation.
	The attributes and values that are observed in the real-world test datasets are highlight in bold.}
	\label{fig:dm}
\end{figure}

Figure~\ref{fig:dm} shows the domain model capturing the test input space
of ADS-DNNs. To develop the domain model, we relied on two sources of information:
\begin{inparaenum}[(1)]
\item the properties that we observed in the real-world ADS-DNN
test datasets (i.e., the Udacity testing datasets~\citep{udacity:dataset}) and
\item the configurable parameters of our simulator.
\end{inparaenum}
In total, we identified four main objects, i.e., \emph{Road}, \emph{Vehicle}, \emph{Weather}, and
\emph{Environment}, and 32 attributes characterizing them, such as \emph{Road.type},
\emph{Vehicle.speed}, \emph{Weather.type}, and \emph{Environment.buildings}.
Each attribute has a specific data type; for example, the \emph{Weather.type}
attribute is an enumeration type, having three different weather values (i.e.,
\emph{Snowy}, \emph{Sunny}, and \emph{Rainy}) as shown in the definition of
\emph{Weather.Type} in Figure~\ref{fig:dm}. This means that only one of the three
values can be assigned to \emph{Weather.type}.
Note that, to illustrate the lower diversity in real-world datasets, the
attributes and their values that are observed in the real world are highlighted
in bold. For example, only the \emph{Sunny} weather is observed in the
real-world test datasets.

In addition to objects and attributes, our domain model includes some
constraints describing valid value assignments to the attributes. These
constraints mostly capture the physical limitations and traffic rules that apply
to our objects. For example, the vehicle speed cannot be higher than 20km/h on
steep curved roads. Constraints may also be used to capture dependencies between
attributes that cannot be specified through the relationships between domain
model objects. For example, we define a constraint to indicate that
\emph{Weather.condition} can only take a value when \emph{Weather.type} is
either \emph{Snowy} or \emph{Rainy}. That is, for \emph{Sunny} we do not need
to specify any weather condition. We have specified these constraints in the
Object Constraint Language (OCL)~\citep{OCL}. The complete OCL constraints are
available in the supporting materials~\citep{supp}.

To produce a simulation scenario (or test scenario) for an ADS-DNN, we
instantiate our domain model in Figure~\ref{fig:dm} by assigning concrete values
to the attributes of our domain model such that its OCL constraints are
satisfied. Specifically, we can represent each test scenario as a vector
$\mathbf{s} = \langle v_1, v_2, \dots, v_{32} \rangle$ where $v_i$ is the value
assigned to the $i$th attribute of our domain model (recall that it contains 32
attributes). We can then initialize the simulator based on the test scenario
vectors. The simulator will then generate, for each of the mobile objects
defined in a scenario, namely the ego and secondary vehicles and pedestrians, a
trajectory vector of the path of that object (i.e., a vector of values
indicating the positions and speeds of the mobile object over time). The length
of the trajectory vector is determined by the duration of the simulation. The
position values are computed based the characteristics of the static objects
specified by the initial configuration, such as roads and sidewalks, as well as
the speed of the mobile objects.

\subsection{Offline Testing}\label{sec:offline}

\begin{figure}
	\centering
	\includegraphics[width=0.7\linewidth]{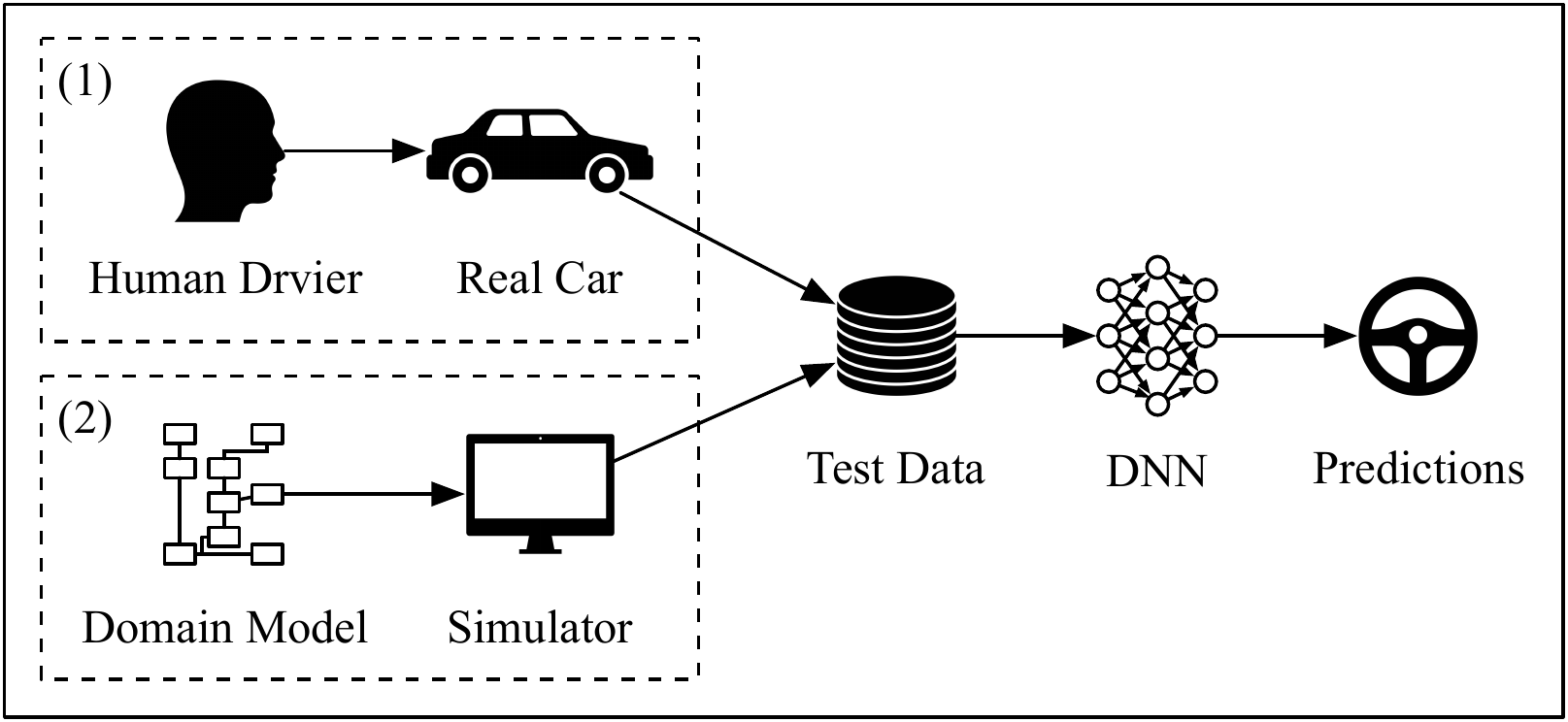}
	\caption{Offline testing using (1) real-world and (2) simulator-generated data}
	\label{fig:offline}
\end{figure}

Figure~\ref{fig:offline} represents an overview of offline DNN testing in the
ADS context. Briefly, offline testing verifies the DNN using historical data
consisting of sequences of images captured from real-life camera or based on a
camera model of a simulator. In either case, the images are labelled with
steering angles. Offline testing measures the \emph{prediction errors} of the
DNN to evaluate test results.

More specifically, let $\mathbf{r}$ be a real-life test dataset composed of a
sequence of tuples $\langle (i^r_1, \theta^r_1), (i^r_2, \theta^r_2), \dots,
(i^r_n, \theta^r_n) \rangle$. For $j=1,\dots,n$, each tuple $(i^r_j,
\theta^r_j)$ of $\mathbf{r}$ consists of an image $i^r_j$ and a steering angle
$\theta^r_j$ label. A DNN $d$, when provided with a sequence $\langle i^r_1,
i^r_2, \dots, i^r_n \rangle$ of the images of $\mathbf{r}$, returns a sequence
$\langle \hat{\theta}^r_1, \hat{\theta}^r_2, \dots, \hat{\theta}^r_n \rangle$ of
predicted steering angles. The prediction error of $d$ for $\mathbf{r}$ is,
then, computed using two well-known metrics, Mean Absolute Error (MAE) and Root
Mean Square Error (RMSE), defined below:

\begin{align*}
\mathit{MAE}(d, \mathbf{r}) &= \frac{\sum_{i=1}^{n} |\theta^r_i - \hat{\theta^r_i}|}{n}
\\
\mathit{RMSE}(d, \mathbf{r}) &= \sqrt{\frac{\sum_{i=1}^{n} (\theta^r_i - \hat{\theta^r_i})^2}{n}}
\end{align*}

To generate a test dataset using a simulator, we provide the simulator with an
initial configuration of a scenario as defined in Section~\ref{sec:domain}. We
denote the offline test dataset generated by a simulator for a scenario
$\mathbf{s}$ by $\mathit{sim}(\mathbf{s}) = \langle (i^s_1, \theta^s_1), (i^s_2,
\theta^s_2), \dots, (i^s_n, \theta^s_n) \rangle$. The prediction error of $d$
for $\mathit{sim}(\mathbf{s})$ is calculated by the MAE and RMSE metrics in the
same way as $\mathit{MAE}(d, \mathbf{r})$ and $\mathit{RMSE}(d, \mathbf{r})$,
replacing $\mathbf{r}$ with $\mathit{sim}(\mathbf{s})$.

\subsection{Online Testing}\label{sec:online}

\begin{figure}
	\centering
	\includegraphics[width=	0.7\linewidth]{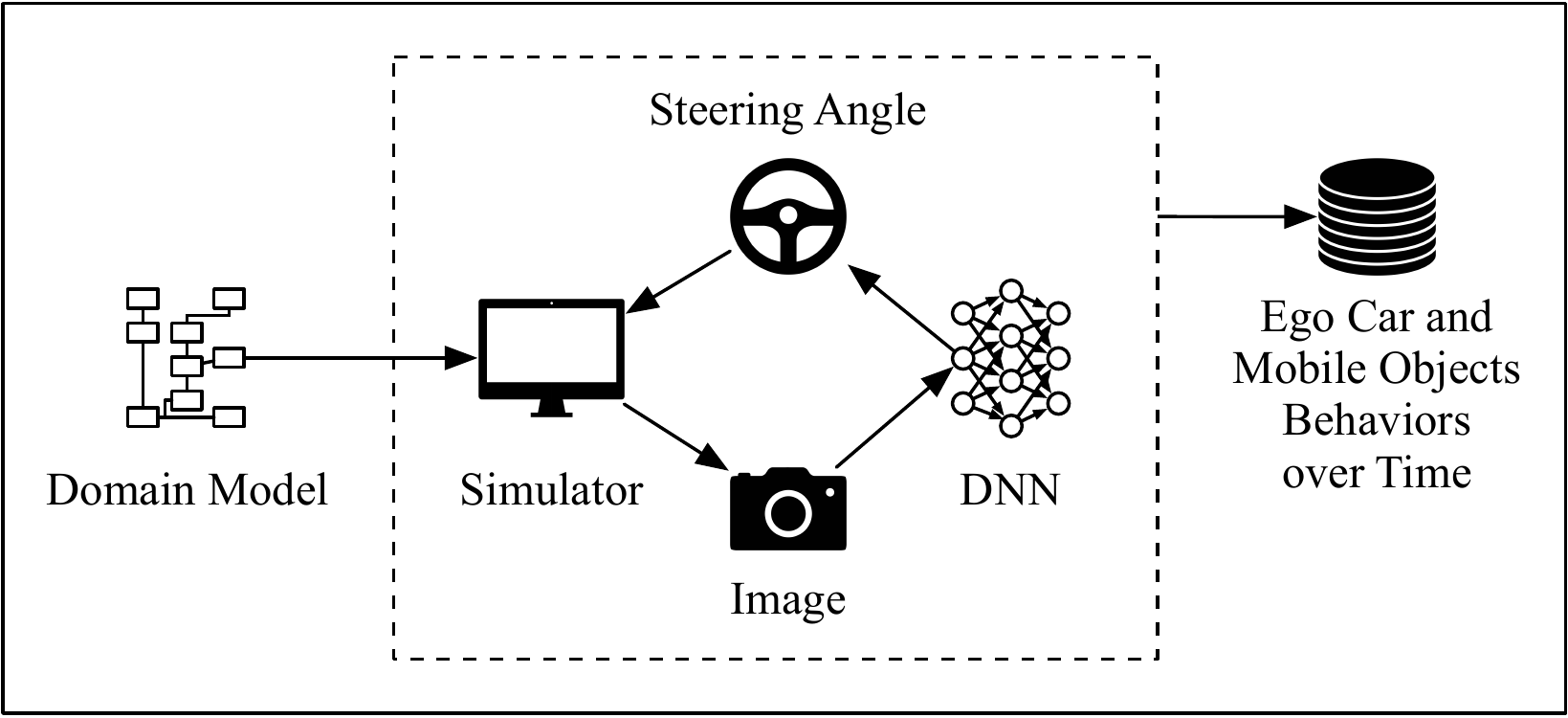}
	\caption{Online testing of ADS-DNNs using simulators}
	\label{fig:online}
\end{figure}

Figure~\ref{fig:online} provides an overview of online testing of DNNs in the
ADS context. In contrast to offline testing, DNNs are embedded into a driving
environment, often in a simulator due to the cost and risk of real-world testing
as we described in Section~\ref{sec:data-sources}. DNNs then receive images
generated by the simulator, and their outputs are directly sent to the (ego)
vehicle models of the simulator. With online testing, we can evaluate how
predictions generated by an ADS-DNN, for an image generated at time $t$ in a
scenario, impact the images to be generated at the time steps after $t$. In
addition to the steering angle outputs directly generated by the ADS-DNN, we
obtain the trajectory outputs of the ego vehicle, which enable us to determine
whether the vehicle is able to stay in its lane.

More specifically, we embed a DNN $d$ into a simulator and run the simulator.
For each (initial configuration of a) scenario, we execute the simulator for a
time duration $T$. The simulator generates the trajectories of mobile objects
as well as images taken from the front-facing camera of an ego vehicle at regular
time steps $t_\delta$, generating outputs as vectors of size $m=\lfloor
\frac{T}{t_\delta} \rfloor$. Each simulator output and image takes an index
between $1$ to $m$. We refer to the indices as simulation time steps. At each
time step $j$, the simulator generates an image $i^s_j$ to be sent to $d$ as
input, and $d$ predicts a steering angle $\hat{\theta}^s_j$ which is
sent to the simulator. The status of the ego vehicle is then updated in the next
time step $j+1$ (i.e., the time duration it takes to update the vehicle is
$t_\delta$) before the next image $i^s_{j+1}$ is generated. In addition to
images, the simulator generates the position of the ego vehicle over time. Recall
that the main function of our DNN is automated lane keeping. This function is
violated when the ego vehicle departs from its lane. To measure the lane departure
degree, we use the Maximum Distance from Center of Lane (MDCL) metric for the
ego vehicle to determine if a safety violation has occurred. The value of MDCL is
computed at the end of the simulation when we have the position vector of the
ego vehicle over time steps, which was guided by our DNN. We cap the value of MDCL
at \SI{1.5}{\meter}, indicating that when MDCL is \SI{1.5}{\meter} or larger,
the ego vehicle has already departed its lane and a safety violation has occurred.
In addition, we normalize the MDCL values between 0 and 1 to make it consistent
with MAE or RMSE.

In this article, we embed the ADS-DNN into PreScan by providing the former with
the outputs from the camera model in input and connecting the steering angle
output of the ADS-DNN to the input command of the vehicle dynamic model.

\section{Experiments}\label{sec:expr}
We aim to compare offline and online testing of DNNs by answering the following research questions:

\textbf{RQ1:} \emph{Can we use simulator-generated data as a reliable
alternative source to real-world data?} Recall the two sources for generating
test data as described in Section~\ref{sec:data-sources}. While
simulator-generated test data is cheaper and faster and is more amenable to
input diversification compared to real-life test data, the texture and
resolution of real-life data look more natural and realistic compared to the
simulator-generated data. In RQ1, we aim to investigate whether, or not, such
differences lead to significant inaccuracies in predictions of the DNN under
test in offline testing. The answer to this question will determine if we can
rely on simulator-generated data for testing DNNs in either offline or online
testing modes.

\textbf{RQ2:} \emph{How frequently do offline and online testing results differ
and do they complement each other?} RQ2 is one of the main research questions we
want to answer in this paper. We want to know how the results obtained by
testing a DNN in isolation, irrespective of a particular application context,
compare with the results obtained by embedding a DNN into a specific
application environment. The answer to this question will help engineers and
researchers better understand the applications and limitations of each testing
mode, and how they could possibly be combined.

\textbf{RQ3:} \emph{Can offline testing results be used to help reduce the cost
of online testing?} In other words, can we focus online testing on situations
where it is needed, i.e., on situations where offline and online testing are in
disagreement? With RQ3, we investigate whether any offline testing results can
be lifted to online testing to help reduce the amount of online testing that we
need to do. Our goal is to determine whether we can characterize the test
scenarios where offline and online testing behave the same in terms of our
domain model elements. This provides the conditions under which offline testing
is sufficient, thus avoiding online testing, which is much more expensive.

\subsection{Experimental Subjects}\label{sec:subjects}

We use three publicly-available, pre-trained DNN-based steering angle prediction
models, i.e., Autumn~\citep{autumn}, Chauffeur~\citep{chauffeur}, and
Kamanda~\citep{komanda}, that have been widely used in previous work to evaluate
various DNN testing approaches~\citep{DeepTest,DeepRoad,Surprise}.

Autumn consists of an image preprocessing module implemented using OpenCV to
compute the optical flow of raw images, and a Convolutional Neural Network (CNN)
implemented using Tensorflow and Keras to predict steering angles. Autumn
improved performance by using cropped images from the bottom half of the entire
images. Chauffeur consists of one CNN that extracts the features from raw images
and a Recurrent Neural Network (RNN) that predicts steering angles from the
previous 100 consecutive images with the aid of a LSTM (Long Short-Term Memory)
module. Similar to Autumn, Chauffeur uses cropped images, and is also
implemented with Tensorflow and Keras. Komanda consists of one CNN followed by
one RNN with LSTM, implemented by Tensorflow, similar to Chauffeur. However, the
underlying CNN of Komanda has one more dimension than Chauffeur that is in
charge of learning spatiotemporal features. Further, unlike Autumn and
Chauffeur, Komanda uses full images to predict steering angles.

\begin{figure}
	\centering
	\includegraphics[width=0.7\linewidth]{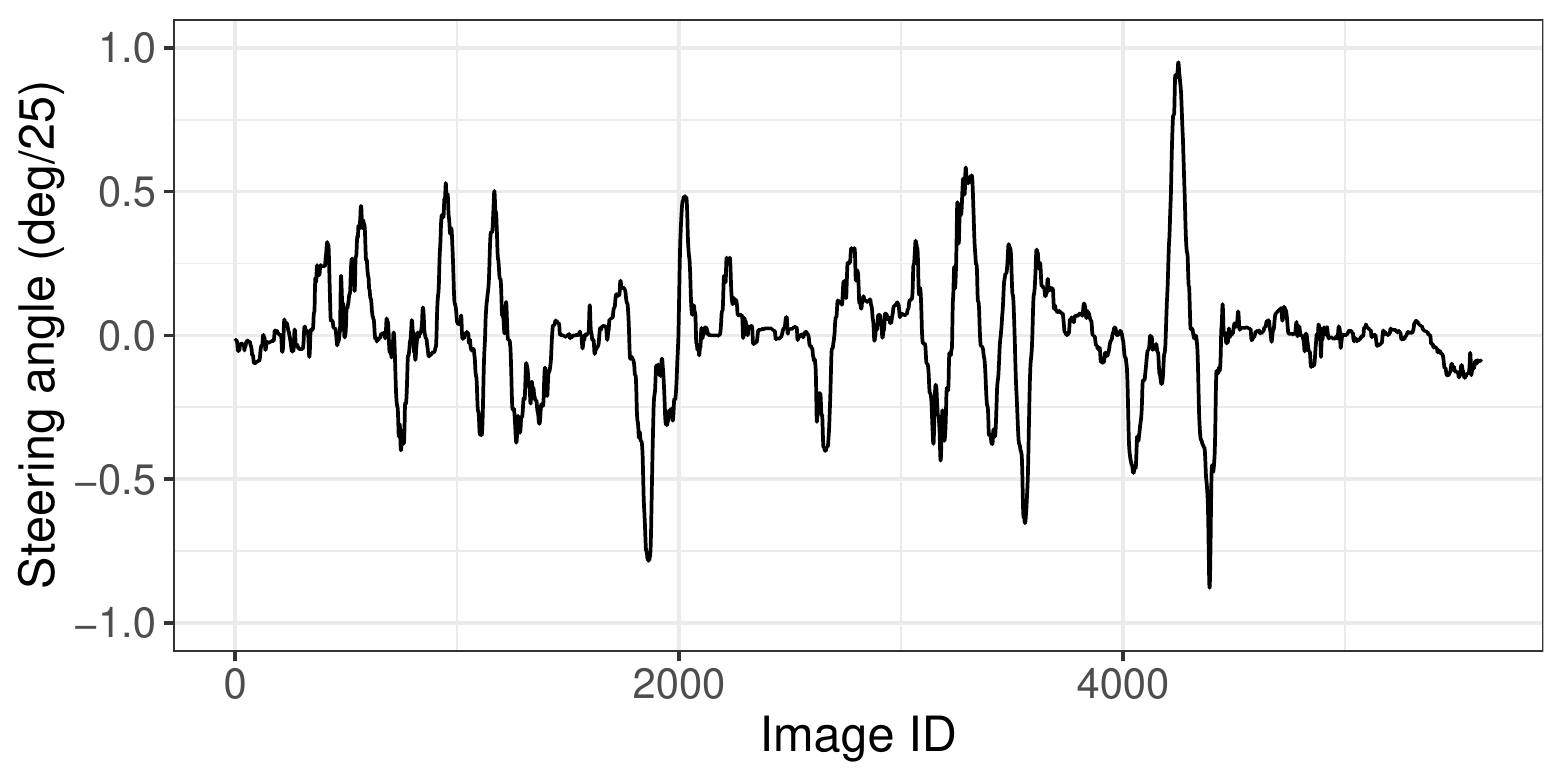}
	\caption{Actual steering angles for the 5614 real-world images used for testing}
	\label{fig:testing-data-stter}
\end{figure}

The models are developed using the Udacity dataset~\citep{udacity:dataset},
which contains 33808 images for training and 5614 images for testing. The images
are sequences of frames of two separate videos, one for training and one for
testing, recorded by a dashboard camera with 20 Frame-Per-Second (FPS). The
dataset also provides, for each image, the actual steering angle produced by a
human driver while the videos were recorded. A positive (+) steering angle
represents turning right, a negative (-) steering angle represents turning left,
and a zero angle represents staying on a straight line. The steering angle
values are normalized (i.e., they are between $-1$ and $+1$) where a $+1$
steering angle value indicates \ang{+25}, and a $-1$ steering angle value
indicates \ang{-25}\footnote{This is how Tian et al.~\citep{DeepTest} have
interpreted the steering angle values provided along with the Udacity dataset,
and we follow their interpretation. We were not able to find any explicit
information about the measurement unit of these values anywhere else.}.
Figure~\ref{fig:testing-data-stter} shows the actual steering angle values for
the sequence of 5614 images in the test dataset. We note that the order of
images in the training and test datasets matters and is accounted for when
applying the DNN models. As shown in the figure, the steering angles issued by
the driver vary considerably over time. The large steering angle values (more
than \ang{3}) indicate actual road curves, while the smaller fluctuations are
due to the natural behavior of the human driver even when the vehicle drives on
a straight road.

Table~\ref{table:rmse} shows the RMSE and MAE values of the two models we
obtained for the Udacity test dataset, as well as the RMSE values reported by
the Udacity website~\citep{udacity:challenge}\footnote{Autumn's RMSE is not
presented in the final leaderboard.}. The differences are attributed to
challenges regarding reproducibility, a well-known problem for state-of-the-art
deep learning methods~\citep{keynote:Joelle} because they involve many
parameters and details whose variations may lead to different results.
Specifically, even though we tried to carefully follow the same settings
and parameters as those suggested on the Udacity website, as shown in
Table~\ref{table:rmse}, the RMSE and MAE values that we computed differed from
those reported by Udacity. We believe these differences are due to the versions
of python and other required libraries (e.g., tensorflow, keras, and scipy). The
precise version information for all these were not reported by Udacity.
Nevertheless, for all of our experiments, we consistently used the most stable
versions of python and the libraries that were compatible with one another.
In other words, irrespective of  differences in the RMSE values between the
reported and our in Table~\ref{table:rmse}, all of our experiments are
internally consistent.
To enable replication of our work, we have made our detailed configurations (e.g.,
python and auxiliary library versions), together with supporting materials,
available online~\citep{supp}.

\begin{table}
\centering
\caption{Accuracies of the subject DNN-based models}
\label{table:rmse}
\begin{tabular}{lrrr}
\toprule
Model & Reported RMSE & Our RMSE & Our MAE \\
\midrule
Autumn		&	 Not Presented & 0.049 & 0.034	\\
Chauffeur	&	 0.058 & 0.092 & 0.055	\\
Komanda		&	 0.048 & 0.058 & 0.039 \\
\bottomrule
\end{tabular}
\end{table}

While MAE and RMSE are two of the most common metrics used to measure prediction
errors for learning models with continuous variable outputs, we mainly use MAE
throughout this article because, in contrast to RMSE, MAE values can be directly
interpreted in terms of individual steering angle values. For example,
$\mathit{MAE}(d, \mathbf{r}) = 1$ means that the average prediction error of $d$
for the images in $\mathbf{r}$ is $1$ (\ang{25}). Since MAE is a more intuitive
metric for our purpose, we will only report MAE values in the remainder of this
article.

\subsection{RQ1: Comparing Offline Testing Results for Real-life Data and Simulator-generated Data}\label{sec:rq1}

\subsubsection{Setup}\label{sec:rq1-setup}
We aim to generate simulator-generated datasets closely mimicking the Udacity
real-life test dataset and verify whether the prediction errors obtained by
applying DNNs to the simulator-generated datasets are comparable with those
obtained for their corresponding real-life ones. As explained in
Section~\ref{sec:subjects}, our real-life test dataset is a sequence of 5614
images labelled with their corresponding actual steering angles. If we could
precisely extract the properties of the environment and the dynamics of the ego
vehicle from the real-life datasets, in terms of initial configuration
parameters of the simulator, we could perhaps generate simulated data closely
resembling the real-life videos. However, extracting information from such
video images to generate inputs of a simulator is not possible.

Instead, we propose a two-step heuristic approach to replicate the real-life
dataset using our simulator. Basically, we steer the simulator to generate a
sequence of images similar to the images in the real-life dataset such that the
steering angles generated by the simulator are close to the steering angle
labels in the real-life dataset.

In the first step, we observe the test dataset and manually identify the
information in the images that correspond to some attribute values in our domain
model described in Section~\ref{sec:domain}. We then create a ``restricted''
domain model by fixing the attribute values in our domain model to the values we
observed in the Udacity test dataset. This enables us to steer the simulator to
resemble the characteristics of the images in the test dataset to the extent
possible. Our restricted domain model includes the attributes and its values
that are highlighted in bold in Figure~\ref{fig:dm}. For example, the restricted
domain model does not include weather conditions other than sunny because the
test dataset has only sunny images. This guarantees that the simulator-generated
images based on the restricted domain model represent sunny scenes only. Using
the restricted domain model, we randomly generate a large number of scenarios
yielding a large number of simulator-generated datasets.

In the second step, we aim to ensure that the datasets generated by the
simulator have similar steering angle labels as the labels in the real-life
dataset. To ensure this, we match the simulator-generated datasets with
(sub)sequences of the Udacity test dataset such that the similarities between
their steering angles are maximized. Note that steering angle is \emph{not} a
configurable attribute in our domain model, and hence, we could not force the
simulator to generate data with steering angle values identical to those in the
test dataset by restricting our domain model. In other words, we minimize the
differences by selecting the closest simulator-generated datasets from a large
pool of randomly generated ones. To do this, we define, below, the notion of
``comparability'' between a real-life dataset and a simulator-generated dataset
in terms of steering angles.

Let $S$ be a set of randomly generated scenarios using the restricted domain
model, and let $\mathbf{r} = \langle (i^r_1, \theta^r_1), \dots, (i^r_k,
\theta^r_k) \rangle$ be the Udacity test dataset where $k=5614$. We denote by
$\mathbf{r}_{(x,l)} = \langle (i^r_{x+1}, \theta^r_{x+1}), \dots, (i^r_{x+l},
\theta^r_{x+l}) \rangle$ a subsequence of $\mathbf{r}$ with length $l$ starting
from index $x+1$ where $x\in \{0, 1, \dots, k\}$. For a given simulator-generated
dataset $\mathit{sim}(\mathbf{s}) = \langle (i^s_1, \theta^s_1), \dots, (i^s_n,
\theta^s_n) \rangle$ corresponding to a scenario $\mathbf{s}\in S$, we compute
$\mathbf{r}_{(x,l)}$ using the following three conditions:
\begin{align}
l &= n \label{eq:1} \\
x &= \argmin_{x} \sum_{j=1}^{l} |\theta^s_j - \theta^r_{x+j}| \label{eq:2} \\
&\frac{\sum_{j=1}^{l} |\theta^s_j - \theta^r_{x+j}|}{l} \le \epsilon \label{eq:3}
\end{align}
where $\argmin_x f(x)$ returns\footnote{If $f$ has multiple points of the
minima, one of them is randomly returned.} $x$ minimizing $f(x)$, and $\epsilon$
is a small threshold on the average steering angle difference between
$\mathit{sim}(\mathbf{s})$ and $\mathbf{r}_{(x,l)}$.
We say datasets $\mathit{sim}(\mathbf{s})$ and $\mathbf{r}_{(x,l)}$ are
\emph{comparable} if and only if $\mathbf{r}_{(x,l)}$ satisfies the three above
conditions (i.e., \ref{eq:1}, \ref{eq:2} and \ref{eq:3}).

Given the above formalization, our approach to replicate the real-life dataset
$\mathbf{r}$ using our simulator can be summarized as follows: In the first
step, we randomly generate a set of many scenarios $S$ based on the reduced
domain model. In the second step, for every scenario $\mathbf{s} \in S$, we
identify a subsequence $\mathbf{r}_{(x,l)}$ from $\mathbf{r}$ such that
$\mathit{sim}(\mathbf{s})$ and $\mathbf{r}_{(x,l)}$ are comparable.

If $\epsilon$ is too large, we may find that $\mathbf{r}_{(x,l)}$ has steering
angles that are too different from those in $\mathit{sim}(\mathbf{s})$. On the
other hand, if $\epsilon$ is too small, we may not be able to find a
$\mathbf{r}_{(x,l)}$ that is comparable to $\mathit{sim}(\mathbf{s})$ for many
scenarios $\mathbf{s} \in S$ randomly generated in the first step. In our
experiments, we select $\epsilon = 0.1$ (\ang{2.5}) since, based on our
preliminary evaluations, we can achieve an optimal balance with this threshold.

For each comparable pair of datasets $\mathit{sim}(\mathbf{s})$ and
$\mathbf{r}_{(x,l)}$, we measure the \emph{prediction error difference} for the
same DNN to compare the datasets. Specifically, we measure $|\mathit{MAE}(d,
\mathit{sim}(\mathbf{s})) - \mathit{MAE}(d, \mathbf{r}_{(x,l)})|$ of a DNN $d$.
Recall that offline testing results for a given DNN $d$ are measured based on
prediction errors in terms of MAE. If $|\mathit{MAE}(d,
\mathit{sim}(\mathbf{s})) - \mathit{MAE}(d, \mathbf{r}_{(x,l)})| \le 0.1$
(meaning \ang{2.5} of average prediction error across all images), we say that
$\mathbf{r}_{(x,l)}$ and $\mathit{sim}(\mathbf{s})$ yield \emph{consistent}
offline testing results for $d$.

We note that the real-life images in the Udacity test dataset are
multicolored or polychromatic. However, our preliminary evaluation confirmed
that the steering predictions of our DNN subjects do not change more than
\ang{0.006} on average when we convert polychromatic images to
monochromatic images in the Udacity test dataset. Hence, we do not attempt to
make the colors of the simulator-generated images similar to that of the
real-life images as color has little impact on the DNN's predictions.

\subsubsection{Results}\label{sec:rq1-results}
\begin{figure}
   \centering
   \begin{subfigure}[b]{\linewidth}
     \centering
     \includegraphics[width=0.7\linewidth]{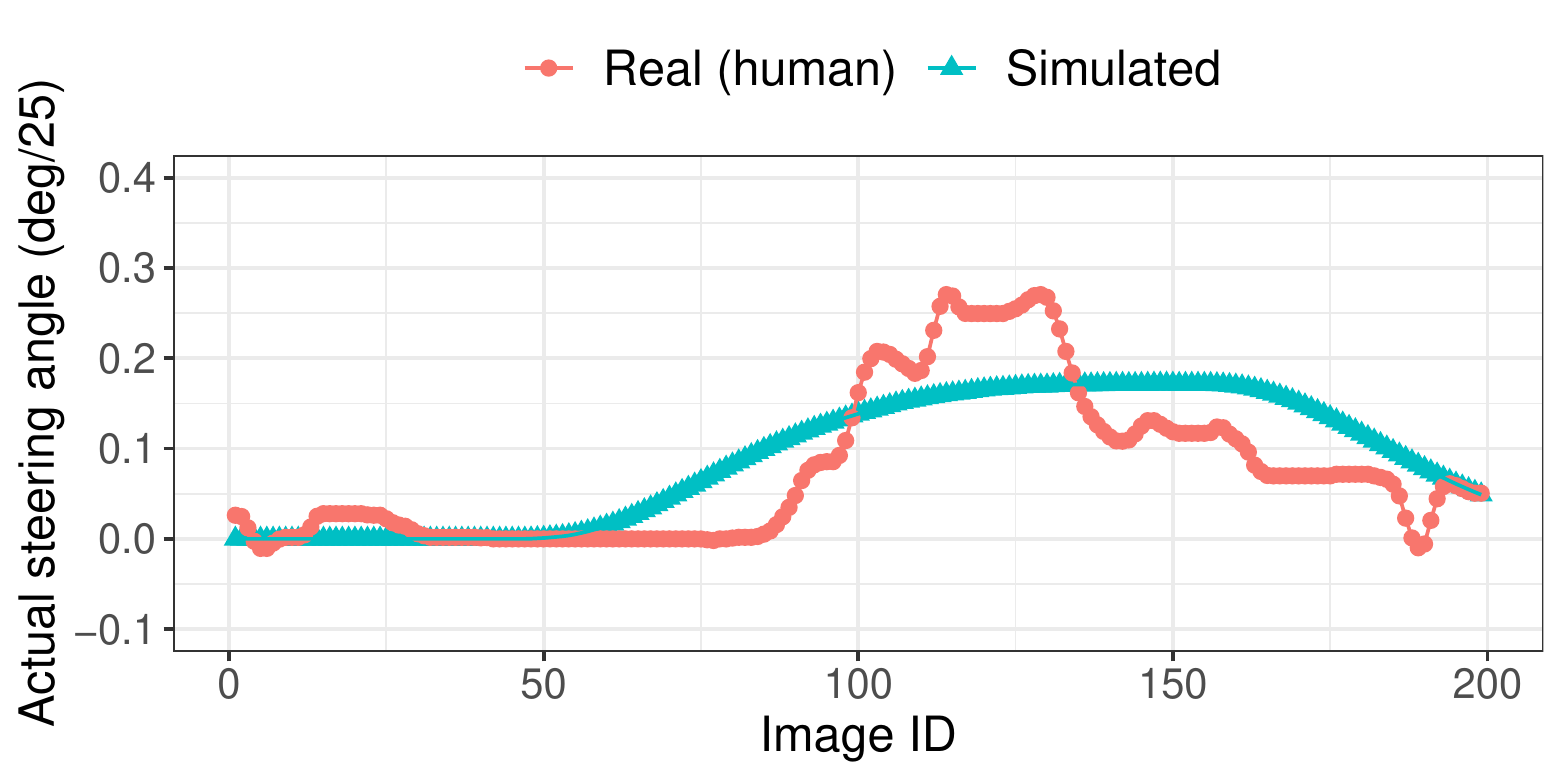}
     \caption{Actual steering angles}
     \label{fig:comparable-steering}
   \end{subfigure}
	\\[3ex]
   \begin{subfigure}[b]{0.35\linewidth}
     \centering
     \includegraphics[width=\linewidth]{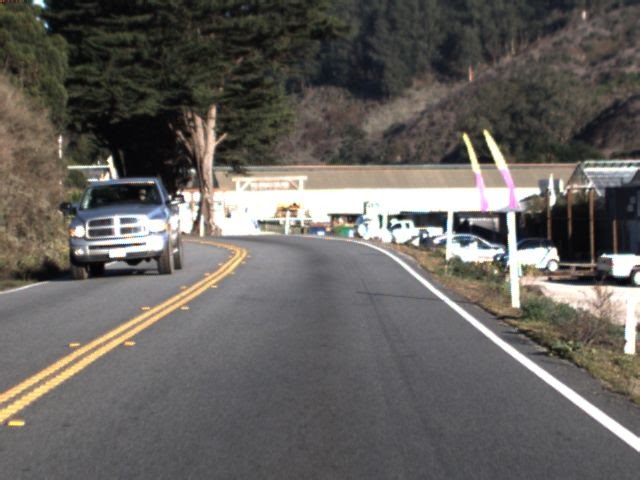}
     \caption{The 40th real image}
     \label{fig:comparable-real1}
   \end{subfigure}
   \begin{subfigure}[b]{0.35\linewidth}
     \centering
     \includegraphics[width=\linewidth]{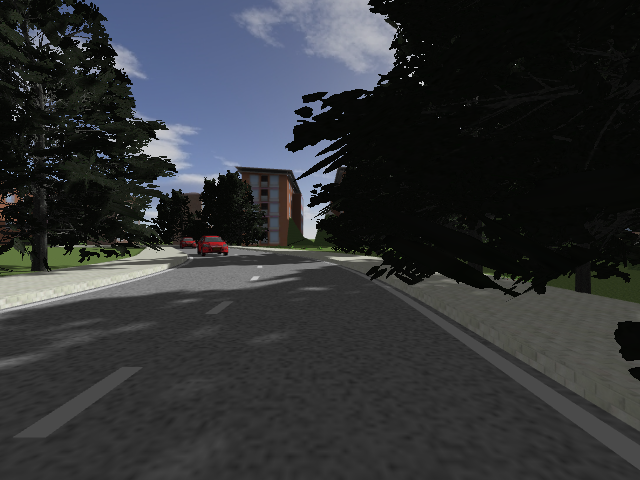}
     \caption{The 40th simulated image}
     \label{fig:comparable-simulated1}
   \end{subfigure}
	 \\[3ex]
	 \begin{subfigure}[b]{0.35\linewidth}
		 \centering
		 \includegraphics[width=\linewidth]{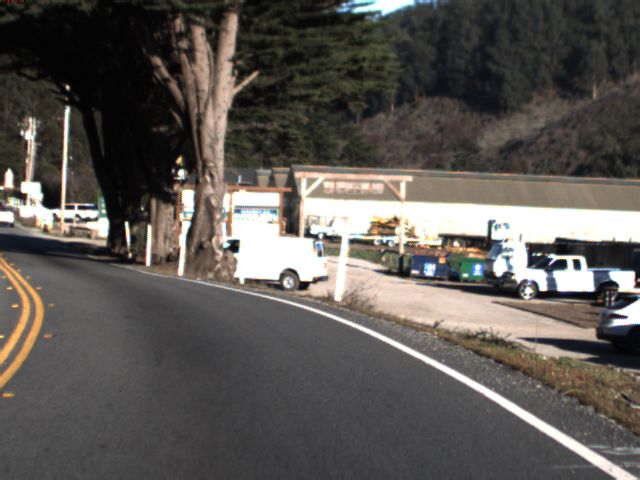}
		 \caption{The 112th real image}
		 \label{fig:comparable-real2}
	 \end{subfigure}
	 \begin{subfigure}[b]{0.35\linewidth}
		 \centering
		 \includegraphics[width=\linewidth]{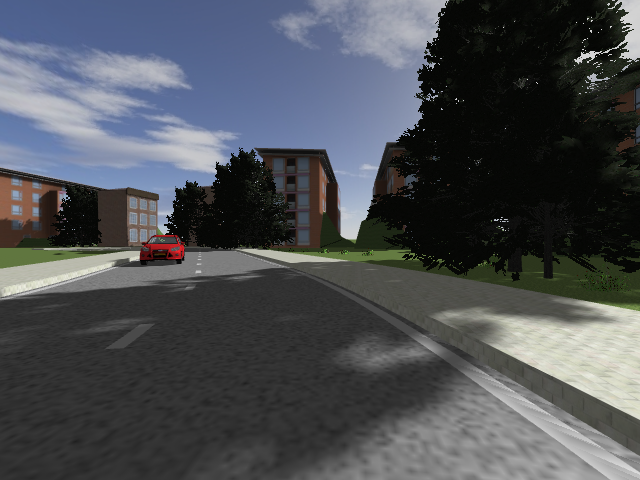}
		 \caption{The 112th simulated image}
		 \label{fig:comparable-simulated2}
	 \end{subfigure}
   \caption{Example comparable pair of a simulator-generated and real-life datasets}
   \label{fig:comparable-pair}
\end{figure}

Among the 100 randomly generated scenarios (i.e., $|S| = 100$), we identified 92
scenarios that could match subsequences of the Udacity real-life test dataset.
Figure~\ref{fig:comparable-pair} shows an example comparable pair of
$\mathbf{r}_{(x,l)}$ (i.e., real dataset) and $\mathit{sim}(\mathbf{s})$ (i.e.,
simulator-generated dataset) identified using our two-step heuristic.
Specifically, Figure~\ref{fig:comparable-steering} shows the steering angles for
all the images in  the example comparable pair.
Figures~\ref{fig:comparable-real1} and \ref{fig:comparable-simulated1} show two
matching frames from the pair where the difference in the steering angles is the
smallest (i.e., the 40th frames where $|\theta^r - \theta^s| = 0$).
Figures~\ref{fig:comparable-real2} and \ref{fig:comparable-simulated2} show two
other matching frames from the pair where the difference in the steering angles
is the largest (i.e., the 112th frames where $|\theta^r - \theta^s| = 0.1115$).
As shown in the steering angle graph in Figure~\ref{fig:comparable-steering},
the simulator-generated dataset and its comparable real dataset subsequence do
not have identical steering angles. For example, the actual steering angles
produced by a human driver have natural fluctuations whereas the steering angles
generated by the simulator are relatively smooth. The differences in steering
angles can also be attributed to the complexity of the real-world not reflected
in the simulator (e.g., bumpy roads). Nevertheless, the overall steering angle
patterns are very similar. If we look at the matching frames shown in
Figures~\ref{fig:comparable-real1} and \ref{fig:comparable-simulated1}, the
matching frames look quite similar in terms of essential properties, such as
road topology and incoming vehicles on the other lane. Regarding the matching
frames shown in Figures~\ref{fig:comparable-real2} and
\ref{fig:comparable-simulated2}, they capture the largest difference in steering
angles of the comparable pair of real and simulated datasets. We can note
differences between the matching frames regarding some aspects, such as the
shape of buildings and trees. Once again, this is because the complexity and
diversity of the real-world is not fully reflected in the simulator. This point
will be further discussed in Section~\ref{sec:dis-online}.

\begin{figure}
	\centering
	\includegraphics[width=0.5\linewidth]{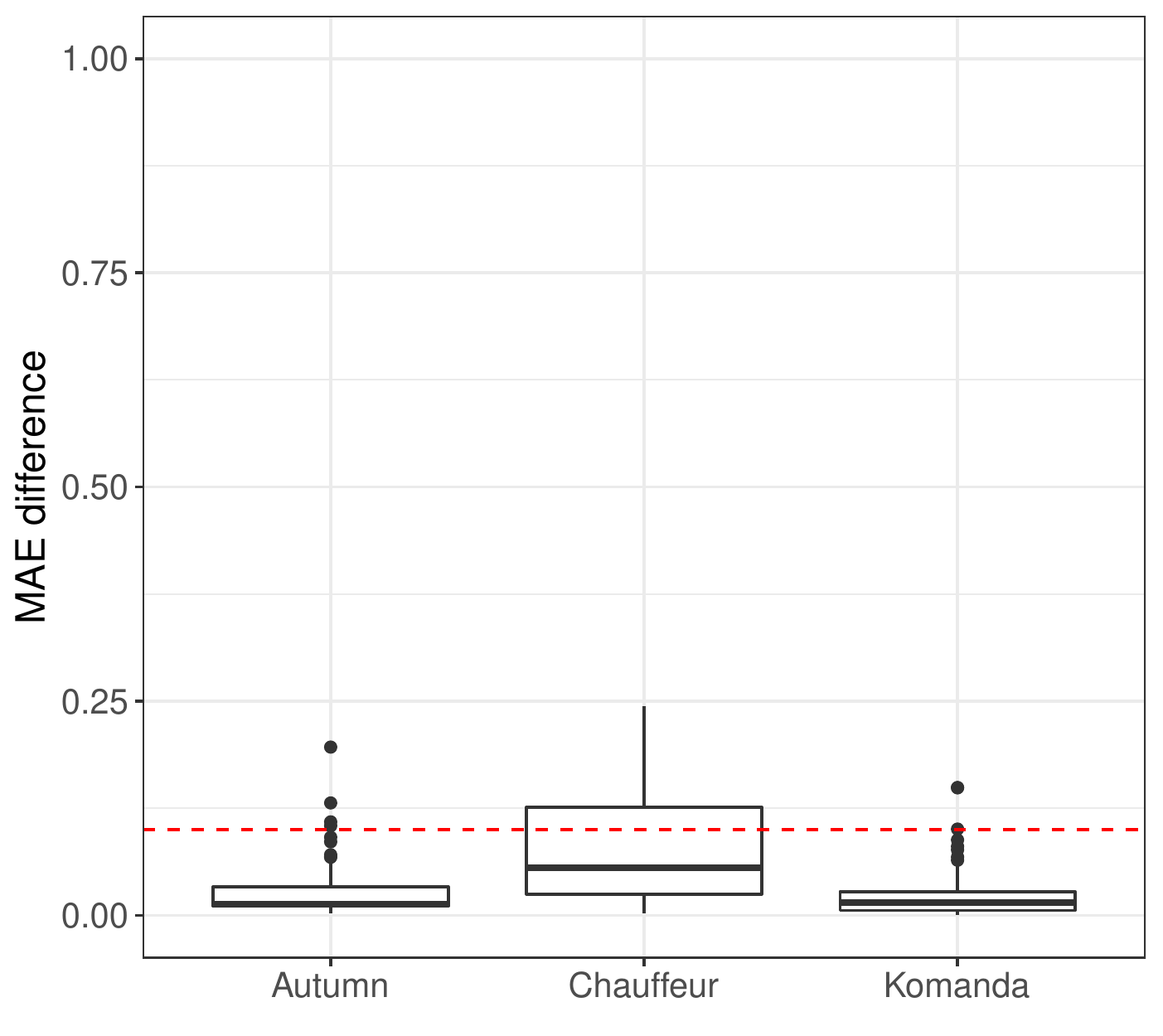}
	\caption{Distributions of the differences between the prediction errors
	obtained for the real datasets (subsequences) and the simulator-generated datasets}
	\label{fig:rq1-boxplot}
\end{figure}

Figure~\ref{fig:rq1-boxplot} shows, for each of our DNNs, Autumn, Chauffeur, and
Komanda, the distributions of the prediction error differences for the real
datasets (subsequences) and the simulator-generated datasets.
For Autumn, the average prediction error difference between the real datasets
and the simulator-generated datasets is $0.027$. Further, 95.6\% of the
comparable pairs show a prediction error difference below 0.1 (\ang{2.5}). This
means that the (offline) testing results obtained for the simulator-generated
datasets are consistent with those obtained using the real-world datasets for
almost all comparable dataset pairs.
The results for Komanda are similar: the average prediction error difference is
$0.023$, and 96.7\% of the comparable pairs show a prediction error difference
below 0.1 (\ang{2.5}).
On the other hand, for Chauffeur, only 66.3\% of the comparable pairs show a
prediction error difference below $0.1$. This means that testing results between
real datasets and simulator-generated datasets are inconsistent in 33.71\% of
the 92 comparable pairs. Specifically, for \emph{all} the inconsistent cases, we
observed that the MAE value for the simulator-generated dataset is greater than
its counterpart for the real-world dataset. It is therefore clear that the
prediction error of Chauffeur tends to be larger for the simulator-generated
dataset than for the real-world dataset. In other words, the simulator-generated
datasets tend to be conservative for Chauffeur and report more false positives
than for Autumn and Komanda in terms of prediction errors. We also found that,
in several cases, Chauffeur's prediction errors are greater than $0.2$ while
Autumn's and Komanda's prediction errors are less than $0.1$ for the same
simulator-generated dataset. One possible explanation is that Chauffeur is
over-fitted to the texture of real images, while Autumn is not thanks to the
image preprocessing module.
Nevertheless, the average prediction error differences between the real datasets
and the simulator-generated datasets is $0.080$ for Chauffeur, which is still
less than $0.1$. This implies that, although Chauffeur will lead to more false
positives (incorrect safety violations) than Autumn and Komanda, the number of
false positives is still unlikely to be overwhelming.

We remark that the choice of simulator as well as the way we generate data
using our selected simulator, based on carefully designed experiments such as
the ones presented here, are of great importance. Selecting a suboptimal
simulator may lead to many false positives (i.e., incorrectly identified
prediction errors) rendering simulator-generated datasets ineffective.

\begin{framed}
\noindent The answer to RQ1 is that, for all the subject DNNs, the prediction error differences between simulator-generated and real-life datasets are less than 0.1 on average. We conclude that we can use simulator-generated datasets as a reliable alternative to real-world datasets for testing DNNs.
\end{framed}

\subsection{RQ2: Comparison between Offline and Online Testing Results}\label{sec:rq2}

\subsubsection{Setup}\label{sec:rq2-setup}
We aim to compare offline and online testing results in this research question.
We randomly generate scenarios and compare the offline and online testing
results for each of the simulator-generated datasets.

For the scenario generation, we use the extended domain model (see
Figure~\ref{fig:dm}) to take advantage of all the feasible attributes provided by
the simulator. Specifically, in Figure~\ref{fig:dm}, the gray-colored entities
and attributes in bold are additionally included in the extended domain model
compared to the restricted domain model used for RQ1. For example, the (full)
domain model contains various weather conditions, such as rain, snow, and fog,
in addition to sunny.

Let $S'$ be the set of randomly generated scenarios based on the (full) domain model.
For each scenario $\mathbf{s}\in S'$, we prepare the simulator-generated dataset
$\mathit{sim}(\mathbf{s})$ for offline testing and measure $\mathit{MAE}(d,
\mathit{sim}(\mathbf{s}))$ for a DNN $d$.
For online testing, we measure $\mathit{MDCL}(d, \mathbf{s})$.
Then we compute the Spearman rank correlation coefficient $\rho$ (rho)
between $\mathit{MAE}(d, \mathit{sim}(\mathbf{s}))$ and $\mathit{MDCL}(d, \mathbf{s})$ to assess the
overall correlation between offline and online testing results. When $\rho$
is 0, it means that there is no monotonic relation between MAE and
MDCL. The closer $\rho$ to 1, the closer the relation between
MAE and MDCL to a perfectly monotonic relation. When $\rho$
is 1, it means that MAE systematically increases (decreases)
when MDCL increases (decreases).

We further compare the offline and online testing results for individual
scenarios. However, since MAE and MDCL are different
metrics, we cannot directly compare them. Instead, we set threshold values for
MAE and MDCL to translate these metrics into binary
results (i.e., \emph{acceptable} versus \emph{unacceptable}) that can be compared. In
particular, we interpret the online testing results of DNN $d$ for a test
scenario $\mathbf{s}$ as acceptable if $\mathit{MDCL}(d, \mathbf{s}) < 0.7$ and unacceptable
otherwise. Note that we have $\mathit{MDCL}(d, \mathbf{s}) < 0.7$ when the departure
from the centre of the lane observed during the simulation of $\mathbf{s}$ is less than
around one meter. Based on domain expert knowledge, such a departure can be considered safe.
We then compute a threshold value for
MAE that is semantically similar to the $0.7$ threshold for
MDCL. To do so, we calculate the steering angle error that leads to
the vehicle deviating from the centre of the lane by one meter. This, however,
depends on the vehicle speed and the time it takes for the vehicle to reach such
deviation. We assume the speed of the vehicle to be 30 km/h (i.e., the
slowest vehicle speed when the vehicle is driving on normal roads) and the time
required to depart from the centre of the lane to be 2.7 seconds (which is a
conservative driver reaction time for braking~\citep{mcgehee2000driver}).
Given these assumptions, we compute the steering angle error corresponding
to a one meter departure to be around \ang{2.5}.
Thus, we consider the offline testing results of $d$ for $\mathbf{s}$
as acceptable if $\mathit{MAE}(d, \mathit{sim}(\mathbf{s})) < 0.1$ (meaning the average
prediction error is less than \ang{2.5}) and unacceptable otherwise.

\subsubsection{Results}\label{sec:rq2-results}
\begin{figure}
     \centering
     \begin{subfigure}[b]{\linewidth}
         \centering
         \includegraphics[width=\linewidth]{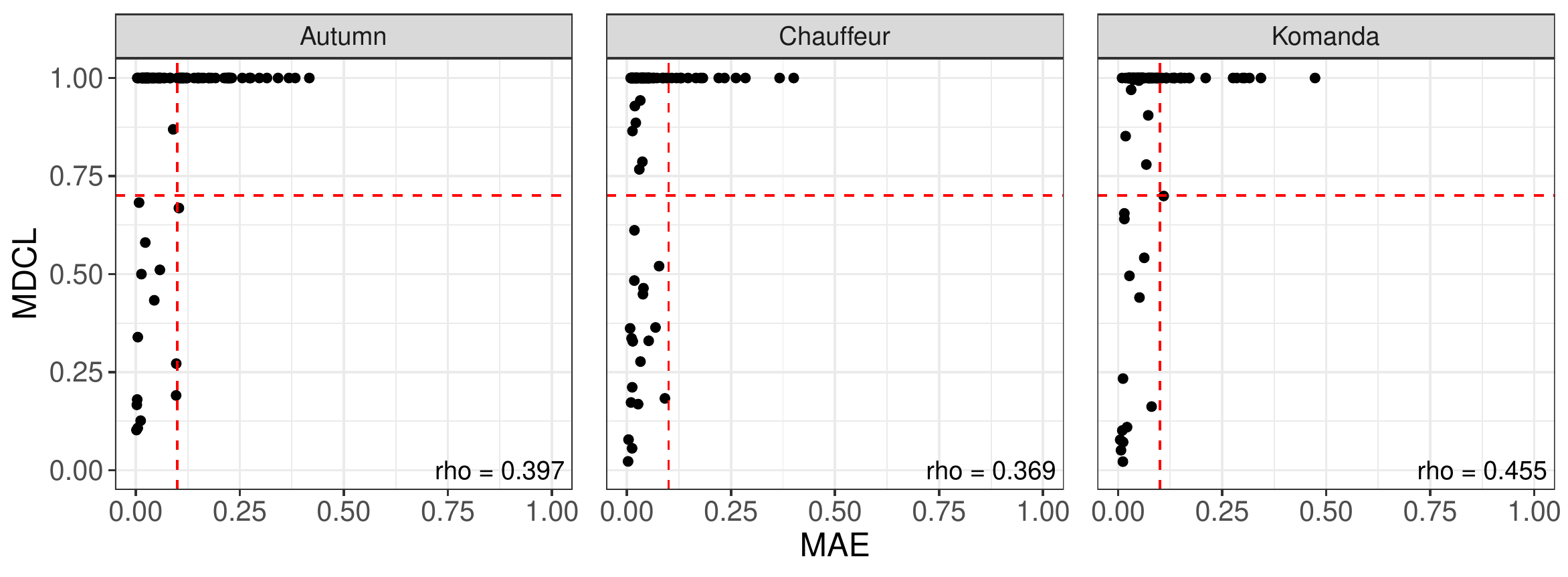}
         \caption{Full view}
         \label{fig:rq2-scatter-overview}
     \end{subfigure}
	\\[3ex]
     \begin{subfigure}[b]{\linewidth}
         \centering
         \includegraphics[width=\linewidth]{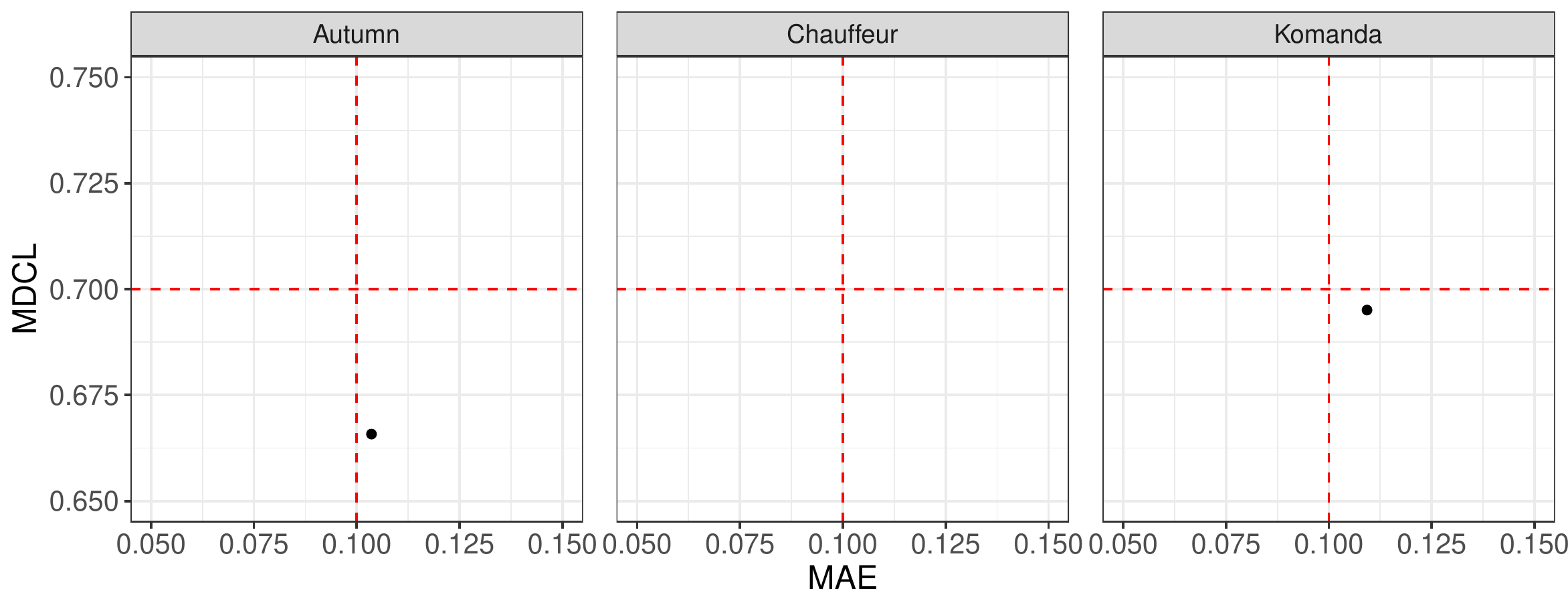}
         \caption{Close-up of the border line}
         \label{fig:rq2-scatter-enlarged}
     \end{subfigure}
     \caption{Comparison between offline and online testing results for all scenarios}
     \label{fig:rq2-scatter}
\end{figure}

Figure~\ref{fig:rq2-scatter} shows the comparison between offline and online
testing results in terms of MAE and MDCL values for all the randomly generated
scenarios in $S'$ where $|S'| = 90$. We generated 90 scenarios because it
is the number of scenarios required to achieve 2-way combinatorial coverage\footnote{We
use PICT (\url{https://github.com/microsoft/pict}) to compute combinatorial coverage.} for
all the attributes in our extended domain model. The x-axis is MAE (offline
testing) and the y-axis is MDCL (online testing). The dashed lines represent the
thresholds, i.e., $0.1$ for MAE and $0.7$ for MDCL. In the bottom-right
corner of each diagram in Figure~\ref{fig:rq2-scatter}, we show the Spearman
correlation coefficients ($\rho$) between MAE and MDCL. For our three DNN
models, $\rho$ is not zero but less than 0.5, meaning that there are weak
correlations between MAE and MDCL.

In Table~\ref{table:contingency}, we have the number of scenarios classified by
the offline and online testing results based on the thresholds. The results show
that offline testing and online testing are not in agreement for 45.5\%, 63.3\%,
and 60.0\% of the 90 randomly generated scenarios for Autumn, Chauffeur,
and Komanda, respectively. Surprisingly, we have only two cases
(one from Autumn and one from Komanda) where the online testing result is
acceptable while the offline testing result is not, and even these two
exceptional cases are very close to the border line as shown in
Figure~\ref{fig:rq2-scatter-enlarged}, i.e., $(0.104, 0.667)$ for Autumn and
$(0.109, 0.695)$ for Komanda where $(x, y)$ indicates MAE=$x$ and MDCL=$y$.
After analyzing the online testing results of these two cases in more detail,
we found that MDCL was less than the threshold simply because the road was short,
and would have been larger had the road been longer.
Consequently, the results show that offline testing is significantly more optimistic than
online testing for the disagreement scenarios.

\begin{table}
\caption{Number of scenarios classified by offline and online testing results}
\label{table:contingency}
\renewcommand{\arraystretch}{1.2}
	\begin{subtable}{\linewidth}
    	\centering
	    \caption{Autumn}
		\begin{tabular}{|r|rr|r|}
			\hline
							& MAE $<$ 0.1	& MAE $\ge$ 0.1	& Total	\\ \hline
			MDCL $<$ 0.7	& 13			& 1				& 14	\\
			MDCL $\ge$ 0.7	& 40			& 36			& 76	\\ \hline
			Total			& 53			& 37			& 90	\\ \hline
		\end{tabular}
    \end{subtable}%
    \vspace*{7pt}
    \begin{subtable}{\linewidth}
	    \centering
        \caption{Chauffeur}
		\begin{tabular}{|r|rr|r|}
			\hline
							& MAE $<$ 0.1	& MAE $\ge$ 0.1	& Total	\\ \hline
			MDCL $<$ 0.7	& 18			& 0				& 18	\\
			MDCL $\ge$ 0.7	& 57			& 15			& 72	\\ \hline
			Total			& 75			& 15			& 90	\\ \hline
		\end{tabular}
    \end{subtable}%
    \vspace*{7pt}
    \begin{subtable}{\linewidth}
	    \centering
        \caption{Komanda}
		\begin{tabular}{|r|rr|r|}
			\hline
							& MAE $<$ 0.1	& MAE $\ge$ 0.1	& Total	\\ \hline
			MDCL $<$ 0.7	& 13			& 1				& 14		\\
			MDCL $\ge$ 0.7	& 53			& 23			& 76	\\ \hline
			Total			& 66			& 24			& 90	\\ \hline
		\end{tabular}
    \end{subtable}
\end{table}

\begin{figure}
     \centering
     \begin{subfigure}[b]{0.35\linewidth}
         \centering
         \includegraphics[width=\linewidth]{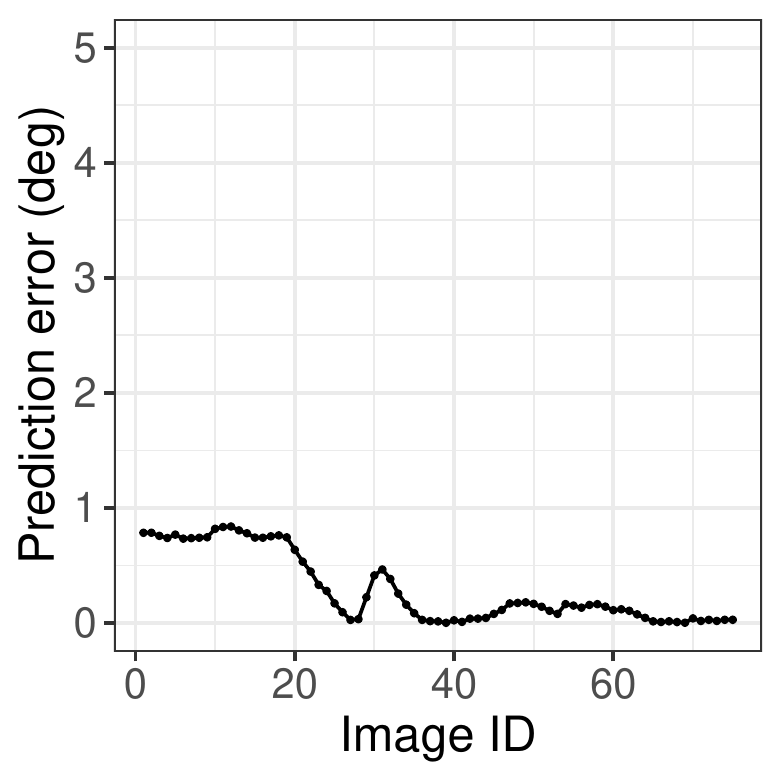}
         \caption{Offline testing result}
         \label{fig:disagree-offline}
     \end{subfigure}
     \begin{subfigure}[b]{0.35\linewidth}
         \centering
         \includegraphics[width=\linewidth]{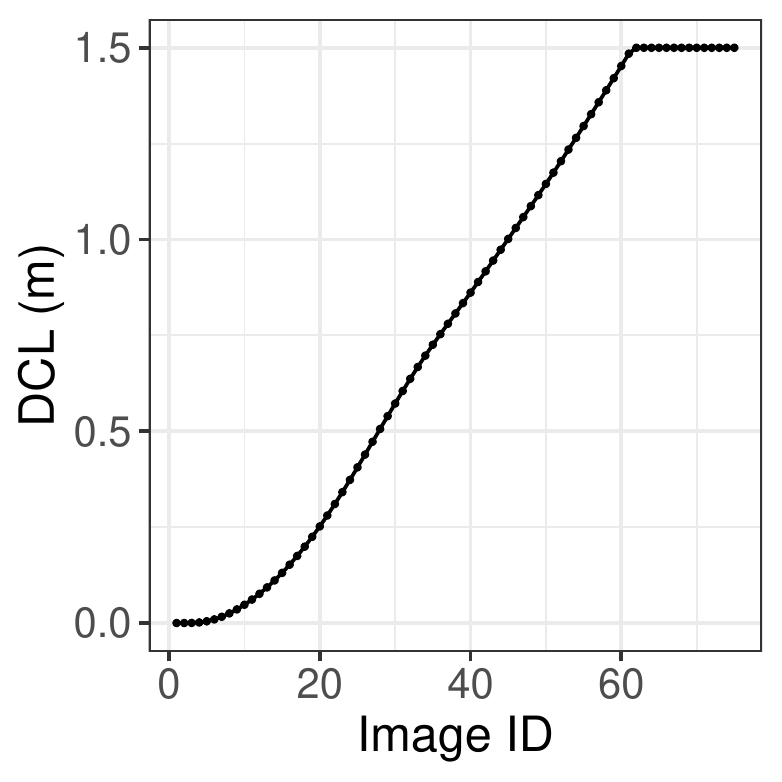}
         \caption{Online testing result}
         \label{fig:disagree-online}
     \end{subfigure}
        \caption{Example inconsistent results between offline and online testing}
        \label{fig:rq2-disagree}
\end{figure}

Figure~\ref{fig:rq2-disagree} shows one of the scenarios on which
offline and online testing disagreed. As shown in Figure~\ref{fig:disagree-offline},
the prediction error of the DNN for each image is always less than \ang{1}.
This means that the DNN appears to be accurate enough according to offline testing.
However, based on the online testing result in Figure~\ref{fig:disagree-online},
the ego vehicle departs from the center of the lane in a critical way
(i.e., more that \SI{1.5}{\meter}). This is because, over
time, small prediction errors accumulate, eventually causing a critical lane
departure. Such accumulation of errors over time is only observable in
online testing, and this also explains why there is no case where the online
testing result is acceptable while the offline testing result is not.

\begin{framed}
\noindent The answer to RQ2 is that offline and online testing results differ in many cases (45.5\%, 63.3\%, and 60.0\% of all scenarios for Autumn, Chauffeur, and Komanda, respectively). We found that offline testing cannot properly reveal safety violations in ADS-DNNs, because it does not account for their closed-loop behavior. Given the fact that detecting safety violations in ADS is the ultimate goal of ADS-DNN testing, we conclude that online testing is preferable to offline testing for ADS-DNNs.
\end{framed}

\subsection{RQ3: Rule Extraction}\label{sec:rq3}

\subsubsection{Setup}\label{sec:rq3-setup}
In RQ2, we showed that, for ADS-DNNs, offline prediction errors are
not correlated with unsafe deviations observed during online testing. In other
words, offline testing will not reveal some of the safety violations that can be
revealed via online testing. However, offline and online testing are both essential
steps in development and verification of DNNs~\citep{9000651}. A typical workflow for
DNN testing is to first apply offline testing, which is a standard Machine
Learning process, and then move to online testing, which is more expensive
and requires engineers to invest significant time on integrating the DNN into a
simulated application environment. The goal of this research question is to
provide guidelines on how to combine offline and online testing
results to increase the effectiveness of our overall testing approach (i.e., to reveal
the most faults) while reducing testing cost. To achieve this goal, we
identify conditions specified in terms of our domain model attributes
(Figure~\ref{fig:dm}) that characterize when offline and online testing agree
and when they disagree. We seek to derive these conditions for different DNN
subjects. Provided with these conditions, we can identify testing scenarios that
should be the focus of online testing, i.e., scenarios for which
offline testing is ineffective, but online testing may reveal a safety violation.

In our work, as discussed in Section~\ref{sec:domain}, each test scenario
is specified as a vector of values assigned to the attributes in our domain
model. By applying offline and online testing to all test scenario
vectors, we can determine if it belongs to the category where offline and online
testing results are in agreement or not. Using test vectors and their
corresponding categories as a set of labelled data instances, we can then
generate classification rules by applying well-known rule mining algorithms,
such as RIPPER~\citep{COHEN1995115}, to learn conditions on domain model
attributes that lead to agreement or disagreement of offline and online testing
results. To be able to learn these conditions with a high degree of accuracy,
however, we need to gather a large collection of labelled data instances
including test vectors ideally covering all combinations of value assignments to
the attributes of our domain model. However, the number of all combinations of
all the attribute values is more than $2^{32}$ since we have 32 attributes of
enumeration types in our domain model that can take more than two values. Furthermore,
the simulation time on a desktop with a 3.6 GHz Intel i9-9900k processor with
32 GB memory and graphic card Nvidia GeForce RTX 2080Ti is up to 20-30
minutes for each scenario depending upon attributes like road length and speed
of ego vehicle. Therefore, it is impossible to cover all the combinations of
value assignments to our domain model attributes.

To be able to learn conditions characterizing agreement and disagreement
between offline and online testing in an effective and efficient way, we propose
a three-step heuristic approach that focuses on learning rules
with statistically high confidence while minimizing the number of test vectors
(i.e., value assignments to domain model attributes) required for learning.
Specifically, we incrementally generate new test vectors to be labelled by
focusing on specific attributes to minimize the amount of data needed for
classification while increasing statistical significance.

\begin{figure}
	\centering
	\includegraphics[width=\linewidth]{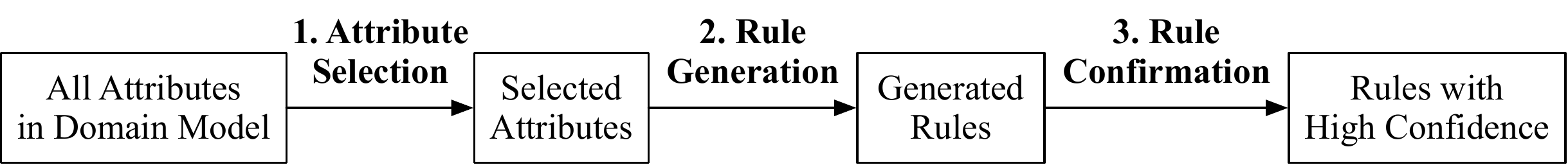}
	\caption{Overall workflow of the three-step heuristic approach to extract rules.}
	\label{fig:rq3-setup}
\end{figure}

Figure~\ref{fig:rq3-setup} outlines the workflow of the approach. The
first step is \emph{attribute selection}, which aims to reduce the search space
by identifying a subset of attributes correlated with the differences between
offline and online testing results for a DNN. The second step (\emph{rule
generation}) aims at extracting rules, for a given DNN, based on the selected
attributes. The third step, \emph{rule confirmation}, seeks to improve the
statistical confidence of the extracted rules. In each of the steps, a minimal
number of new data instances are incrementally generated. The details of the
steps are described next:
\begin{enumerate}
\setlength{\itemsep}{5pt}
\setlength{\parskip}{2pt}
\item \emph{Attribute Selection}: In data mining, attribute selection
(a.k.a., feature selection) strategies often rely on generating a large number
of labelled data instances randomly to ensure data diversity and the uniformity
of their distribution in the search space.
Since labelling data instances (i.e., test vectors) in our work is expensive, we are
limited regarding how many test vectors we can generate and label for the purpose of
attribute selection. Therefore, instead of using a pure random strategy, as the
input for our attribute selection strategy, we use $n$-way combinatorial testing
to generate a relatively small number of diverse test vectors. The value of $n$
is determined based on our time budget for labelling data (running test vectors
in our work) and the number of attributes in our domain. In our work, for the
purpose of attribute selection, we set $n=2$. This led to generating 90 test
vectors to be able to cover the pairwise combination of values of the $32$
attributes in our domain model. For each test vector, we determine whether
offline and online testing results are in agreement or not, resulting in the binary
classification of our test vectors. We then perform the attribute selection
using the Random Forest algorithm~\citep{GENUER20102225}. We use the concept of
variable importance in Random Forests to select important attributes in our domain model
since it has been reported to be accurate in general~\citep{ARCHER20082249,GENUER20102225}.
Table~\ref{table:RQ3-selected-attributes} shows the selected attributes for each
DNN. Six, four, and two attributes are selected for Autumn, Chauffeur, and
Komanda, respectively.

\item \emph{Rule Generation}: At this step, we trim the set of test
vectors generated in the attribute selection step by hiding, from each vector,
the attributes that were not selected in the previous step. The result is a set
of labelled test vectors that only include values for the important attributes
of our domain (i.e., those attributes selected in the previous step). This set,
however, does not necessarily cover different combinations of the values for the
selected attributes. Hence, we enhance this set by generating a number of new
test vectors. We use $n$-way combinatorial test generation again, but this time
we only consider the attributes selected in the previous step in the test
generation (i.e., the attributes that were not selected in step one are simply set a
random value). For this step, we choose $n=3$ since we are dealing with a smaller
number attributes and are able to generate more value combinations during the
same test time budget. For Komanda, we use all combinations
since we have only two selected attributes.
The number of new test vectors therefore varies depending on the
selected attributes for each DNN. The new test vectors, together with the 90 test
vectors generated in the previous step, are used for rule generation. We extract
rules using the RIPPER algorithm~\citep{COHEN1995115}, yielding a set of rules
for our DNN under analysis. Each generated rule is a tuple $(\mathit{if},
\mathit{label})$ where $\mathit{if}$ describes conditions on the values of the
selected attributes (e.g., \emph{Vehicle.speed} $> 10$ $\wedge$ \emph{Road.type} = \emph{Curved}) and
$\mathit{label}$ describes a class (i.e., \emph{agree} or \emph{disagree}).
We can estimate the accuracy of each rule as the number of test vectors labelled by
$\mathit{label}$ and satisfying the $\mathit{if}$ conditions over the number of
test vectors satisfying $\mathit{if}$.
Table~\ref{table:RQ3-generated-rules} shows the rules generated for each DNN. As
shown in the table, we obtain three rules for Autumn, four rules for Chauffeur,
and two rules for Komanda. Each rule has an $\mathit{if}$ part, described as a
conjunction of predicates defined over our domain model attributes, and a
$\mathit{label}$ part that can be either \emph{agree} or \emph{disagree}.
For the accuracy values, we also report the 95\% Confidence Interval (CI). For
example, for the second rule of Autumn, the accuracy value $0.88 \pm 0.20$ means
that the true accuracy value has a 95\% probability of being in the interval [$0.68$  $1.00$].
The CI range is quite large since, in our work, we have minimized the total
number of test vectors, and therefore the number of test vectors satisfying the conditions
$\mathit{if}$ for each rule can be small. Hence we may not be able
establish reasonably narrow CIs for the accuracy values of the generated rules.
To alleviate this issue, we use a third step to increase the statistical
confidence in the estimated accuracy of the generated rules.

\item \emph{Rule Confirmation}: The basic idea is to reduce the CI length
by providing more data instances for each rule. For each rule, we repeatedly
generate a data instance satisfying the $\mathit{if}$ part of the rule until the
CI length of the estimated accuracy of the rule with a 95\% confidence level is less
than a threshold $\lambda$. In our experiments, we set $\lambda = 0.2$, meaning
$\pm$0.1. Note that we generate new data instances only for the extracted rules
to efficiently reduce the CI length of the rules. The final results after
performing the Rule Confirmation step is shown in Table~\ref{table:RQ3-results}
and will be discussed in the next section (Section~\ref{subsec-Rq3results}).
\end{enumerate}

\begin{table}
\caption{Intermediate Results: Selected Attributes}
\label{table:RQ3-selected-attributes}
\centering
\begin{tabular}{ll}
\toprule
DNN       & Selected Attributes                                                                                         \\
\midrule
Autumn    & \emph{Road.type}, \emph{Road.laneLineColor}, \emph{Road.curbLinePattern}, \\ & \emph{Vehicle.laneNumber}, \emph{Vehicle.headLights}, \emph{Weather.condition} \\
Chauffeur & \emph{Road.type}, \emph{Road.roadSpecificProperty}, \emph{Vehicle.fogLights}, \\ & \emph{Environment.underlay}                                                  \\
Komanda   & \emph{Weather.type}, \emph{Environment.bulidings}                \\
\bottomrule
\end{tabular}
\end{table}

\begin{table}
\caption{Intermediate Results: Generated Rules}
\label{table:RQ3-generated-rules}
\centering
\begin{tabularx}{\linewidth}{lXr@{\hskip 3pt}l}
\toprule
DNN                        & Rule                                                                 & \multicolumn{2}{c}{Accuracy} \\
\midrule
\multirow{3}{*}{Autumn}    & If \emph{Road.curbLanePattern} = \emph{Dashed} then \emph{disagree}                & 0.58        & $\pm$ 0.11     \\
                           & If \emph{Road.type} = \emph{Curved} then \emph{disagree}                            & 0.88        & $\pm$ 0.20      \\
                           & Other than mentioned above then \emph{agree}                            & 0.65        & $\pm$ 0.11        \\
\midrule
\multirow{4}{*}{Chauffeur} & If \emph{Vehicle.fogLights} = \emph{True}, then \emph{agree}                               & 0.59        & $\pm$ 0.13         \\
                           & If \emph{Road.type} = \emph{Straight} $\wedge$ \emph{Environment.underlay} = \emph{Pavement} then \emph{agree}         & 0.90        & $\pm$ 0.18      \\
                           & If \emph{Road.type} = \emph{Curved} then \emph{agree}                                 & 0.70        & $\pm$ 0.28    \\
                           & Other than mentioned above then \emph{disagree}                             & 0.76        & $\pm$ 0.09        \\
\midrule
\multirow{2}{*}{Komanda}   & If \emph{Weather.type} = \emph{Rainy} $\wedge$ \emph{Environment.buildings} = \emph{False} then \emph{agree} & 0.76        & $\pm$ 0.18         \\
                           & Other than mentioned above then \emph{disagree}                          & 0.69       & $\pm$ 0.11        \\
\bottomrule
\end{tabularx}
\end{table}

\subsubsection{Results}\label{sec:rq3-results}
\label{subsec-Rq3results}
\begin{table}
\caption{Rule Extraction Results}
\label{table:RQ3-results}
\centering
\begin{tabularx}{\linewidth}{llXr@{\hskip 3pt}l}
\toprule
DNN                        & ID & Rule                                                                 & \multicolumn{2}{c}{Accuracy} \\
\midrule
\multirow{3}{*}{Autumn}    & A1 & If \emph{Road.curbLanePattern} = \emph{Dashed} then \emph{disagree}                    & 0.61        & $\pm$ 0.10       \\
                           & A2 & If \emph{Road.type} = \emph{Curved} then \emph{disagree}                                 & 0.95        & $\pm$ 0.10        \\
                           & A3 & Other than mentioned above then \emph{agree}                                & 0.61        & $\pm$ 0.10       \\
\midrule
\multirow{4}{*}{Chauffeur} & C1 & If \emph{Vehicle.fogLights} = \emph{True}, then \emph{agree}                                   & 0.58        & $\pm$ 0.10       \\
                           & C2 & If \emph{Road.type} = \emph{Straight} $\wedge$ \emph{Environment.underlay} = \emph{Pavement} then \emph{agree}         & 0.71        & $\pm$ 0.10        \\
                           & C3 & If \emph{Road.type} = \emph{Curved} then \emph{agree}                                    & 0.55        & $\pm$ 0.10        \\
                           & C4 & Other than mentioned above then \emph{disagree}                             & 0.76        & $\pm$ 0.09        \\
\midrule
\multirow{2}{*}{Komanda}   & K1 & If \emph{Weather.type} = \emph{Rainy} $\wedge$ \emph{Environment.buildings} = \emph{False} then \emph{agree} & 0.59        & $\pm$ 0.10       \\
                           & K2 & Other than mentioned above then \emph{disagree}                             & 0.67        & $\pm$ 0.10        \\
\bottomrule
\end{tabularx}
\end{table}

Table~\ref{table:RQ3-results} shows the rules generated for our three DNN
subjects after applying the process described in Section~\ref{sec:rq3-setup} and
Figure~\ref{fig:rq3-setup}. Note that in contrast to the results reported in
Table~\ref{table:RQ3-generated-rules}, the accuracy values in
Table~\ref{table:RQ3-results} are those obtained after applying the rule
confirmation step. For example, for the first rule for Autumn, the accuracy of
$0.61 \pm 0.10$ means that around 61\% of the scenarios that satisfy the
condition \emph{Road.curbLanePattern} = \emph{Dashed} are labelled with
\emph{disagree}. Thanks to our rule confirmation step, we are able to ascertain
the accuracy levels of rules within a narrower 95\% confidence interval. In our
work, the rule accuracy indicates the predictive power of the rule. For example,
the second rule for Autumn is highly accurate and hence predictive (more than
85\% of scenarios). Hereafter, for simplicity, we use the IDs indicated in
Table~\ref{table:RQ3-results} to refer to the rules.

Overall, there is no rule predicting \emph{agree} with an accuracy above
0.71. This means that there is no condition with accuracy above 0.71 where
offline testing results conform to online testing results. That is, the test
results for scenarios that match the  ``agree'' rule conditions in
Table~\ref{table:RQ3-results} may still differ during offline and online testing
with a high probability. Therefore, based on our results, we are not able to
identify conditions that can characterize, with a high accuracy, agreement
between offline and online testing to help lift offline testing results to
online testing and reduce the amount of online testing needed. Our results,
further, suggest that, at least for ADS-DNNs, we may not be able to find rules
that can, in general, differentiate between offline and online testing
behaviors. As can be seen from the table, there is not much similarity between
the rules we have obtained for different DNNs. This is because these DNNs have
different architectures, use different features of the input images for
prediction and are trained differently.

However, our observations show that these rules still may provide valuable
insights as to how different DNNs work. When an attribute appears in a rule, it
indicates that the attribute has a significant impact on the DNN output, and
hence, this attribute can be used to classify both the situations where offline
testing is as good as online testing (i.e., DNN prediction errors indeed
indicate a safety violation) as well as the dual situations where offline
testing is simply too optimistic. For example, the attributes
\emph{Weather.type} and \emph{Environment.buildings} appear only in the
conditions for the rules of Komanda. On the other hand, for Chauffeur and
Autumn, the attributes appearing in the rule conditions are related to the road
shape and the road lane patterns.  This confirms the fact that Komanda uses full
images to predict steering angles while the Chauffeur and Autumn focus on the
road-side views in the images, as noted in Section~\ref{sec:subjects}.

Another reason explaining differences across DNNs is that some  DNNs are
inaccurate for certain attributes regardless of testing modes, which  means that
both offline testing and online testing are capable of detecting the  faulty
behaviors of these DNNs with a relatively high probability.  For example, we
found that Chauffeur is, in general, inaccurate for predicting steering  angles
for curved roads. Due to this weakness, both offline and online testing  results
are in agreement when \emph{Road.type} is \emph{Curved}, as shown in C3.

The last reason for differences is that, as shown in RQ1, Chauffeur works
relatively better on real-world images than on simulated images. Since Chauffeur
is not effective with simulated images, it may yield more prediction errors in
offline testing, and hence, offline and online testing results are more likely
to be in agreement as ``unacceptable''. This explains why we have three
``agree'' rules, namely C1, C2 and C3, for Chauffeur while for other DNNs we
have fewer ``agree'' rules.

\begin{framed}
\noindent The answer to RQ3 is that offline testing results cannot be used to reduce the cost of online testing as we are not able to identify conditions that characterize, with a high accuracy, agreement between offline and online testing for all our subject DNNs.
\end{framed}

\subsection{Threats to Validity}\label{sec:threads}
In RQ1, we propose a two-step approach that builds simulator-generated datasets
comparable to a given real-life dataset. While it achieves its objective, as
shown in Section~\ref{sec:rq1-results}, the simulated images are still different
from the real images. However, we confirmed that the prediction errors obtained
by applying our subject DNNs to the simulator-generated datasets are comparable
with those obtained for their corresponding real-life datasets. Thus, the
conclusion that offline and online testing results often disagree with each other is valid.

We used a few thresholds that may affect the experimental results
in RQ2 and RQ3. To reduce the chances of misinterpreting the
results, we selected intuitive and physically interpretable metrics to
evaluate both offline and online test results (i.e, prediction errors and
safety violations), and defined threshold values based on common sense and
experience. Further, adopting different threshold values, as long as they are
within a reasonable range, does not change our findings. For example, if we use
$\mathit{MAE(d, \mathit{sim}(\mathbf{s}))} < 0.05$ as a threshold in offline testing
results instead of $\mathit{MAE(d, \mathit{sim}(\mathbf{s}))} < 0.1$, the numbers of
scenarios in Table~\ref{table:contingency} change. However, it does not change
the correlation analysis results and the fact that we have many scenarios for
which offline and online testing results disagree, nor does it change the
conclusion that offline testing is more optimistic than online testing.

Different ADS-DNNs may lead to different results. For example, we may
able to identify conditions that can characterize, with a high accuracy,
agreement between offline and online testing results to lift offline testing
results to online testing and reduce the amount of online testing needed for a
specific ADS-DNN. To mitigate such a threat, we tried our best to find all
candidate ADS-DNNs in the literature and selected the three subject
ADS-DNNs (i.e., Autumn, Chauffeur, and Komanda) that are publicly available and
sufficiently accurate for steering angle predictions.

Though we focused, in our case study, on only two lane-keeping DNNs (steering
prediction)---which have rather simple structures and do not support braking or
acceleration, our findings are applicable to all DNNs in an ADS context as
long as the closed-loop behavior of the ADS matters.

\section{Discussion}\label{sec:discussion}

\subsection{Online Testing using Simulators}\label{sec:dis-online}
One important purpose of online testing is to test a trained ADS-DNN with
the newest unseen data that potentially appear in the application environments
of the ADS-DNN. However, because simulators cannot express all the complexity
and diversity of the real world, online testing using simulators cannot cover
all possible scenarios in the real world. For example, in the case of online
testing using a simulator that cannot express weather changes, certain safety
violations of the ADS-DNN that occurs only in rainy weather cannot be found.

Nevertheless, considering the problems of online testing in the real world,
especially the cost and risk, online testing using a simulator is inevitable.
Indeed, according to our industrial partners in the automotive industry, due to
the excessive amount of manpower and resources required to collect real-world
data, it is impossible to gather sufficient and diverse real-world data. On the
other hand, a simulator can generate sufficiently diverse data at a much lower
cost and risk.

Furthermore, simulator-based test input generation has an additional advantage
regarding the test oracle problem~\citep{6963470}. When we use simulators for
online testing, the generation of test oracles is completely automated. For
real-life datasets, however, test oracles may need to be manually specified
which is labor-intensive and time-consuming. For example, for ADS-DNNs, the
driver's maneuvers and the data gathered from the various sensors and cameras
during online testing in the real world may not contain sufficient information
to automatically generate test oracles. In contrast, simulators are able to
generate labeled datasets, from which test oracles can be automated, for various
controlling, sensing, and image recognition applications. But, as expected, the
accuracy of test results and oracles depends on the fidelity of simulators.

\subsection{Offline vs. Online Testing: What to Use in Practice?}\label{sec:dis-comp}
Experimental results show that offline testing cannot properly detect
safety requirements violations identified in online testing. Offline testing is
in fact inadequate to identify faulty behaviors for ADS-DNNs with closed-loop
behavior. In other words, online testing is essential to adequately detect
safety violations in ADS-DNNs, where interactions with the application
environment are important. In particular, online testing using a simulator is
highly recommended if a high-fidelity simulator is available.

However, online testing is not essential in all cases. When testing
ADS-DNNs without closed-loop behavior, offline and online testing results are
expected to be similar because errors are not accumulating over time. For
example, in the case of an ADS-DNN that simply warns the driver instead of
directly controlling the steering when necessary, there is no closed-loop since
the DNN's predictions do not actually control the vehicle, and therefore offline
testing would be sufficient.

\subsection{Open Challenges}\label{sec:dis-chal}
There are also challenges that need to be addressed in online testing. For
example, the higher the fidelity of a simulator, the more time it takes to
simulate, which has a direct impact on the cost of online testing. In particular,
when using a search-based technique, online testing may take a very long time
because the simulator must be repeatedly executed for various scenarios.
Therefore, more research is required to reduce the cost of online testing.

Research on high-fidelity simulators is also essential. As discussed in
Section~\ref{sec:dis-online}, online testing using a simulator cannot completely
cover all possible scenarios in the real world. However, by utilizing a
simulator higher fidelity, the risk of uncovered scenarios could be
significantly reduced~\citep{airsim18,9294422}. Research on how to lower the
risk through a more systematic approach is also needed.

\section{Related Work}\label{sec:offline-online}
Table~\ref{table:related-papers} summarizes DNN testing approaches specifically
proposed in the context of autonomous driving systems. Approaches to the
general problem of testing machine learning systems are discussed in the recent
survey by \citet{9000651}.

\begin{sidewaystable}
\centering
\caption{Summary of DNN testing studies in the context of autonomous driving}
\label{table:related-papers}
\renewcommand{\arraystretch}{1.3}
\begin{tabularx}{\linewidth}{p{3cm}lp{2.3cm}X}
\toprule
Author(s) & Testing mode & DNN's role & Summary \\
\midrule
\citet{dreossi2017systematic} & Offline & Object detection & Test image generation by arranging basic objects using greedy search  \\
\citet{DeepXplore} & Offline & Lane keeping & Coverage-based label-preserving test image generation using joint optimization with gradient ascent \\
\rowcolor{lightgray}
\citet{Codevilla_2018_ECCV} & Offline and online & Lane keeping & Improving the correlation between offline and online testing results by selecting an appropriate testing dataset and suitable offline metrics \\
\citet{DeepTest} & Offline & Lane keeping & Coverage-based label-preserving test image generation using greedy search with simple image transformations \\
\rowcolor{lightgray}
\citet{8500421} & Online & Object detection & Test scenario generation using the combination of covering arrays and simulated annealing \\
\citet{89960-2_22} & Offline & Traffic sign recognition & Adversarial image generation using feature extraction \\
\citet{DeepRoad} & Offline & Lane keeping & Label-preserving test image generation using Generative Adversarial Networks (GANs) \\
\citet{zhou2018deepbillboard} & Offline & Lane keeping & Adversarial billboard-image generation for digital and physical adversarial perturbation \\
\rowcolor{lightgray}
\citet{Gambi2019issta} & Online & Lane keeping & Automatic virtual road network generation using search-based Procedural Content Generation (PCG) \\
\citet{Surprise} & Offline & Lane keeping & Improving the accuracy of DNNs against adversarial examples using surprise adequacy \\
\rowcolor{lightgray}
\citet{majumdar2019paracosm} & Online & Object detection, \newline lane keeping & Test scenario description language and simulation-based test scenario generation to cover parameterized environments \\
\citet{Zhou:2019:MTD} & Offline & Object detection & Combination of Metamorphic Testing (MT) and fuzzing for 3-dimensional point cloud data \\
\midrule
\rowcolor{lightgray}
This article & Offline and online & Lane keeping & Comparison between offline and online testing results and investigate if we can use offline testing results to run fewer tests during online testing \\
\bottomrule
\end{tabularx}
\end{sidewaystable}

In Table~\ref{table:related-papers}, approaches for online testing are
highlighted grey. As the table shows, most of existing approaches focus on the
offline testing mode only, where DNNs are seen as individual units without accounting for the
closed-loop behavior of a DNN-based ADS. Their goal is to generate test data
(either images or 3-dimensional point clouds) that lead to DNN prediction errors.
\citet{dreossi2017systematic} synthesized images for driving
scenes by arranging basic objects (e.g., road backgrounds and vehicles) and
tuning image parameters (e.g., brightness, contrast, and saturation).
\citet{DeepXplore} proposed \textsc{DeepXplore}, an approach that
synthesizes images by solving a joint optimization problem that maximizes both
neuron coverage (i.e., the rate of activated neurons) and differential behaviors
of multiple DNNs for the synthesized images.
\citet{DeepTest} presented \textsc{DeepTest}, an approach that
generates label-preserving images from training data using greedy search for
combining simple image transformations (e.g., rotate, scale, and for and rain
effects) to increase neuron coverage.
\citet{89960-2_22} generated adversarial examples, i.e., small
perturbations that are almost imperceptible by humans but causing DNN
misclassifications, using feature extraction from images.
\citet{DeepRoad} presented \textsc{DeepRoad}, an approach that
produces various driving scenes and weather conditions by applying Generative
Adversarial Networks (GANs) along with corresponding real-world weather scenes.
\citet{Zhou:2019:MTD} combined Metamorphic Testing (MT) and Fuzzing
for 3-dimensional point cloud data generated by a LiDAR sensor to reveal
erroneous behaviors of an object detection DNN.
\citet{zhou2018deepbillboard} proposed \textsc{DeepBillboard}, an
approach that produces both digital and physical adversarial billboard images to
continuously mislead the DNN across dashboard camera frames. While this work is
different from the other offline testing studies as it introduces adversarial
attacks through sequences of frames, its goal is still the generation of test
images to reveal DNN prediction errors.
In contrast, \citet{Surprise} defined a coverage criterion, called
\emph{surprise adequacy}, based on the behavior of DNN-based systems with
respect to their training data. Images generated by \textsc{DeepTest} were
sampled to improve such coverage and used to increase the accuracy of the DNN
against adversarial examples.

Online testing studies exercise the ADS closed-loop behavior and generate test
driving scenarios that cause safety violations, such as unintended lane
departure or collision with pedestrians.
\citet{8500421} were the first to raise the problem that previous
works mostly focused on the DNNs, without accounting for the closed-loop
behavior of the system.
\citet{Gambi2019issta} also pointed out that testing DNNs for ADS
using only single frames cannot be used to evaluate closed-loop properties of
ADS. They presented \textsc{AsFault}, a tool that generates virtual roads which
cause self-driving cars to depart from their lane.
\citet{majumdar2019paracosm} presented a language for describing
test driving scenarios in a parametric way and provided \textsc{Paracosm}, a
simulation-based testing tool that generates a set of test parameters in such a
way as to achieve diversity.
We should note that all the online testing studies rely on virtual (simulated)
environments, since, as mentioned before, testing DNNs for ADS in real traffic
is dangerous and expensive. Further, there is a growing body of evidence
indicating that simulation-based testing is effective at finding violations.
For example, recent studies for robotic applications show that simulation-based
testing of robot function models not only reveals most bugs identified  during
outdoor robot testing, but that it can additionally reveal several bugs that
could not have been detected by outdoor testing~\citep{8009918}.

There is only one study comparing offline and online testing results by
investigating the correlations between offline and online testing prediction
error metrics~\citep{Codevilla_2018_ECCV}. The authors found that the
correlation between offline prediction and online performance is weak, which is
consistent with the results of this article. They also found two ways for improving
the correlations: (1) augmenting the testing data (e.g., include images from
three cameras, i.e., a forward-facing one and two lateral cameras facing 30
degrees left and right, instead of having images from one forward-facing camera)
and (2) selecting a proper offline testing metric (e.g., Mean Absolute Error
other than Mean Squared Error). Their analysis relies on the offline and online
testing of DNNs trained by simulator-generated images, while our DNNs
are trained with real-world images. Nevertheless, consistent with our results, they
concluded that offline testing is not adequate. Furthermore, beyond simple
correlations and in order to draw more actionable conclusions, our investigation
looked at whether offline testing was a sufficiently reliable mechanism for
detecting safety violations in comparison to online testing. Last, we
investigated whether offline and online testing results could agree under
certain conditions, so as to take advantage of the lower cost of offline testing
in such situations.

\section{Conclusion}\label{sec:conclusion}

This article presents a comprehensive case study to compare two distinct testing
phases of Deep Neural Networks (DNNs), namely offline testing and online
testing, in the context of Automated Driving Systems (ADS). Offline testing
evaluates DNN prediction errors based on test data that are generated
without involving the DNN under test. In contrast, online testing
determines safety requirement  violations of a DNN-based system in a specific
application environment based on  test data generated dynamically from
interactions between the DNN under test and  its environment. We aimed to
determine \emph{how offline and online testing  results differ or complement
each other} and \emph{if we can exploit offline testing  results to run
fewer tests during online testing to reduce the testing cost}. We additionally
investigated if we can use simulator-generated datasets as a  reliable
substitute to real-world datasets for DNN testing.

The experimental results on the three best performing ADS-DNNs from the Udacity
Self-Driving Car Challenge 2~\citep{udacity:challenge} show that
simulator-generated datasets yield DNN prediction errors that are similar to
those obtained by testing DNNs with real-world datasets. Also, offline testing
is more optimistic than online testing as many safety violations identified by
online testing could not be identified by offline testing, while large
prediction errors generated by offline testing always led to severe safety
violations detectable by online testing. Furthermore, the experimental results
show that we cannot exploit offline testing results to reduce the
cost of online testing in practice since we are not able to identify specific
situations where offline testing could be as accurate as online testing in
identifying safety violations.

The results of this paper have important practical implications for DNN testing,
not only in an ADS context but also in other CPS where the closed-loop behavior
of DNNs matters. Specifically, both researchers and practitioners should focus
more on online testing as offline testing is not able to properly determine
safety requirement violations of the DNN-based systems under test.

Considering the expensive cost of online testing, even using a high-fidelity
simulator instead of a real-world environment, our results also call for more
efficient online testing approaches. As part of future work, we plan to develop
an approach for automatic test scenario generation using surrogate models and
search-based testing to efficiently identify safety critical test scenarios for
online testing.

\bibliographystyle{spbasic}      %
\bibliography{ADS-DNN-Testing}   %

\end{document}